\documentclass[11pt]{article}
\pdfoutput=1
\usepackage[margin=1.0in]{geometry}
\usepackage{amsmath,amssymb,graphicx}
\usepackage{hyperref}
\usepackage{slashed}

\usepackage[T1]{fontenc}

\usepackage[utf8]{inputenc}

\usepackage[greek,english]{babel}

\usepackage[font=small,labelfont=bf]{caption}
\usepackage{cite}

\newcommand{\iu}{{\mathrm i}}

\newcommand{\RR}{{\mathbb{R}}}
\newcommand{\ZZ}{{\mathbb{Z}}}

\newcommand{\be}{\begin{equation}}
\newcommand{\ee}{\end{equation}}

\usepackage{xcolor}
\usepackage{bm}

\title{\bf \huge  Reflections on Parity Breaking}
\author{Jacob McNamara$^{1}$ and Matthew Reece$^{2}$\\
{\small \color{gray} \texttt{jmcnamar~(@caltech.edu), mreece~(@g.harvard.edu)}}\\
{\small $^{1}$Walter Burke Institute for Theoretical Physics}\\
{\small California Institute of Technology, Pasadena, CA 91125, USA}\\
{\small $^{2}$Department of Physics, Harvard University, Cambridge, MA, 02138, USA}}

\begin{document}
\maketitle

\begin{abstract}
Parity and CP symmetries are broken in the world around us. Nonetheless, parity (or CP) may be a gauge symmetry which is higgsed in our universe. This is assumed in many scenarios for physics beyond the Standard Model, including the classic Nelson--Barr proposal for the Strong CP problem. Gauged parity can only arise in quantum gravity, where it corresponds to a path integral over both orientable and non-orientable spacetime manifolds. We show that spontaneous breaking of gauged parity leads to exactly stable domain walls, and describe the implications for the cosmology of models with gauged parity. These domain walls carry an unusual sort of charge, which superficially has features in common with both gauge charges and global charges. We show that these unusual charges are consistent with the expected absence of global symmetries in quantum gravity when there exists a complete spectrum of dynamical objects required by the Swampland Cobordism Conjecture, including end-of-the-world branes.
\end{abstract}

\tableofcontents

\section{Introduction}

Parity violation~\cite{Lee:1956qn, Wu:1957my} and CP violation~\cite{Landau:1957tp, Lee:1957qq, Christenson:1964fg} are established facts about the world around us. In the Standard Model, parity is not a symmetry in any sense, being incompatible with the field content (choices of gauge representations) of the theory. CP, on the other hand, is a possible symmetry given the field content, which is observed to be violated by the CKM phase $\delta_\textsc{CKM}$~\cite{Kobayashi:1973fv,Jarlskog:1985ht}. It is intriguing that, so far, CP violation has been observed only in conjunction with flavor violation in the quark sector (and perhaps the neutrino sector as well, with $\delta_\textsc{PMNS}$ measurements from T2K and NO$\nu$A in some tension at the time of this writing~\cite{T2K:2021xwb, NOvA:2021nfi}). Meanwhile, CP violation {\em without} flavor violation is stringently constrained by null results in searches for the neutron EDM~\cite{Abel:2020pzs} (giving rise to the Strong CP problem) and the electron EDM~\cite{ACME:2018yjb} (sensitive to multi-TeV scale physics beyond the Standard Model). 

The peculiar status of CP violation, appearing in some observables and not in others, has given rise to the idea that CP (or some other generalized parity symmetry) may be a spontaneously broken symmetry of nature. This idea plays a role in a number of phenomenological applications, perhaps most prominently in the approach to the Strong CP problem pioneered by Nelson~\cite{Nelson:1983zb} and Barr~\cite{Barr:1984qx}. There is a large body of work along such lines; a sampling of classic and recent references includes~\cite{Beg:1978mt, Mohapatra:1978fy, Babu:1989rb, Barr:1991qx, Bento:1991ez, Hiller:2001qg, Dine:2015jga, Albaid:2015axa, Hall:2018let, Craig:2020bnv, deVries:2021pzl}. Along different lines, Nir and Rattazzi suggested that if CP is spontaneously broken by flavon fields, which also carry charges under horizontal flavor symmetries, this could accommodate TeV-scale new physics while explaining the absence of flavor-conserving but CP-violating signals in EDMs~\cite{Nir:1996am}, an idea that has  recently been extended to the lepton sector in light of current data~\cite{Aloni:2021wzk, Nakai:2021mha}. 

Although the idea of spontaneous CP violation has been studied extensively in particle phenomenology, little of this work has confronted basic conceptual questions: what would it mean for parity (or CP) to be an exact symmetry of nature? What are the consequences of its breaking, including in the early universe?  Exceptions include two prominent papers that discussed the possibility of CP as a gauge symmetry~\cite{Dine:1992ya, Choi:1992xp}, emphasizing the possible origin of 4d CP in spacetime symmetries of ten-dimensional superstring theory, as in a classic  example of Strominger and Witten~\cite{Strominger:1985it}. Whereas~\cite{Dine:1992ya} indicated  that CP domain walls would need to be inflated  away,~\cite{Choi:1992xp} suggested a dynamical process that could destroy the domain wall. In this paper, we argue that such a process is topologically impossible: CP domain walls are absolutely stable. 

A common pitfall in discussions of parity symmetry is that parity is introduced as a specific symmetry of Minkowski space: $(t, \vec x) \mapsto (t, -\vec x)$.\footnote{A remark is in order about terminology. Some references use the term ``parity'' for the reversal of all spatial coordinates, which is orientation-reversing for an odd number of spatial dimensions, but is orientation-preserving (part of the connected rotation group) in an even number of spatial dimensions.  In this context one often speaks not of parity but of R or reflection symmetry, which reverses a single spacetime coordinate, and correspondingly of CR or CRT in place of CP or CPT. Since our primary interest is in 4d spacetime, we use the convention that ``parity'' refers to any orientation-reversing element of the full Lorentz group $O(1, d-1)$, with the understanding that in even numbers of spatial dimensions this does {\em not} include the operation that reverses all spatial coordinates.} Though this definition is sufficient for non-gravitational physics in flat space, it is not immediately clear how to generalize to curved spacetime backgrounds (a necessity, to discuss gravity and cosmology). While there are special cases of spacetime backgrounds which admit orientation-reversing isometries, which could play the role of parity symmetry, there is no general reason to expect spacetime to have such a global isometry even for a theory that locally admits parity symmetry.

The proper resolution is well-known in the literature on formal QFT and on condensed matter theory (CMT), but has been mostly absent from the particle physics literature. A theory with an exact parity symmetry is one that can be defined on non-orientable manifolds. After traversing an orientation-reversing cycle in the manifold, a particle returns as its parity conjugate. This definition of parity can be readily extended to more general orientation-reversing spacetime symmetries like CP, where the spacetime parity transformation is combined with an internal symmetry when traversing an orientation-reversing cycle.\footnote{In fact, there is no invariant distinction between P and CP symmetry, and the two symmetries may be exchanged by dualities (see, e.g.,~\cite{Cordova:2017kue}). Thus, it is a matter of convention which to call P, and which to call CP.} One can view a QFT defined on a non-orientable manifold as a QFT in the presence of a ``background parity gauge field'' induced by the parity-reversing transition functions of the manifold.

When a QFT may be placed on non-orientable manifolds in a consistent way, parity is a global symmetry, which may be gauged~\cite[Footnote 1]{Witten:2016cio}. To gauge a symmetry, we must sum over background gauge fields in the path integral. As the background parity gauge field is built from the transition functions of the spacetime manifold, it only makes sense to speak of parity as a gauge symmetry in the context of quantum gravity, where we sum over spacetime manifolds in the path integral. For a theory with gauged parity, this sum is over both orientable and non-orientable manifolds.\footnote{For example, parity is a global symmetry of the two dimensional theory of gravity on the Type IIB string worldsheet, while parity is gauged on the Type I string worldsheet.} For an alternative argument that gauged parity only makes sense in quantum gravity, recall that the full Lorentz group is a semidirect product,
\begin{equation}
O(1, d-1) = SO(1, d-1) \rtimes \ZZ_2^{\rm P},
\end{equation}
of the proper Lorentz group with parity symmetry $\ZZ_2^{\rm P}$. Because $\ZZ_2^{\rm P}$ is not a normal subgroup of $O(1,d-1)$, there is no way to both gauge parity and preserve Lorentz symmetry without gauging the full Lorentz group, which is only possible in a gravitational theory. UV complete, holographic theories of quantum gravity with parity symmetry must have gauged parity~\cite{Harlow:2018tng}.
 
There has been substantial investigation of orientation-reversing spacetime symmetries in the literature on formal QFT and CMT. For example, it is known that such symmetries can have 't~Hooft anomalies, e.g., the classic ``parity anomaly'' in 3d theories~\cite{Redlich:1983dv, Niemi:1983rq, Alvarez-Gaume:1984zst} or more recent mixed 't~Hooft anomalies involving time-reversal and 1-form symmetries in 4d gauge theories~\cite{Gaiotto:2017yup}. Time reversal symmetries also play a key role in classifying possible phases in CMT, e.g.,~\cite{Kane:2005zz, Schnyder:2008tya, Chen:2012ctz, Tachikawa:2016cha}. In the context of quantum gravity, M-theory has been studied on non-orientable manifolds~\cite{Witten:2016cio, Freed:2019sco}. Recently, non-invertible time reversal symmetries have been found in gauge theories~\cite{Choi:2022rfe}.

Our goal in this paper is to connect the formal understanding of parity in terms of physics on non-orientable manifolds with the particle physics question of whether CP is a spontaneously broken symmetry of the world around us. In particular, we argue that spontaneous CP breaking leads to exactly stable domain walls, as a consequence of simple topology. The exact stability of parity domain walls is the main take-away message of this paper for particle phenomenologists and cosmologists. It is a potentially severe problem for the cosmology of Nelson--Barr models or other phenomenological scenarios involving spontaneously broken CP. 

Our results are also of interest for formal study of UV complete quantum gravity, in which symmetries, including parity, are expected to be either broken or gauged~\cite{Zeldovich:1976vq, Banks:1988yz, Banks:2010zn, Harlow:2018tng}. Familiar characteristic classes in mathematics provide potential symmetries in QFT and quantum gravity~\cite{McNamara:2019rup, Heidenreich:2020pkc}. An alternative statement of our main result is that in quantum gravity theories with gauged parity symmetry, the would-be $(d-2)$-form symmetry arising from the first Stiefel-Whitney class $w_1$ is always gauged.

\medskip

The paper is organized as follows. We begin in \S\ref{sec:discretegauge} with a discussion of discrete {\em internal} global and gauge symmetries. We explain why domain walls associated with the spontaneous breaking of global symmetries are stable, whereas the domain walls for spontaneously broken gauge symmetries can end on vortices (cosmic strings, in 4d), rendering the domain walls unstable. This section is entirely a review of familiar results, and could be skipped by readers who are comfortable with this material, though it does connect with very recent work on generalized symmetries and the Completeness Hypothesis \cite{Rudelius:2020orz, Heidenreich:2021xpr, McNamara:2021cuo, Casini:2020rgj, Casini:2021zgr, Henning:2021ctv, Roumpedakis:2022aik, Bhardwaj:2022yxj, Arias-Tamargo:2022nlf, Bhardwaj:2022scy, Bartsch:2022mpm}.

After this pedagogical review, the next two sections contain our main results of interest to a particle phenomenology audience. We discuss parity and its spontaneous breaking in~\S\ref{sec:parityDW}. We give a proof of the stability of parity domain walls in QFT with parity as a global symmetry, which completely parallels the case of internal global symmetries. This establishes that any decay of a parity domain wall is a process that changes the topology of spacetime itself, and hence necessarily a quantum gravity process. We explain the relationship between orientability, the first Stiefel-Whitney class, and parity domain walls. Our story then diverges from the case of an internal symmetry: we show that parity domain walls remain absolutely stable even once parity is gauged in a gravitational theory. In~\S\ref{sec:models}, we sketch the cosmological implications of the stability of parity domain walls, and discuss some model-building options.

The stability of parity domain walls raises the question: what symmetry protects them? We answer this question in the following two sections, which rely upon more mathematics than the earlier sections.  In \S\ref{sec:gauge_charge}, we argue that parity domain walls carry a type of generalized $(d-2)$-form gauge charge, albeit an unusual one that at first sight has features in common with both gauge charges and global charges. In~\S\ref{sec:cobordism}, we interpret our results in terms of the Swampland Cobordism Conjecture \cite{McNamara:2019rup} and the Adams Spectral Sequence. We discuss other dynamical objects that can exist in gravitational theories with gauged parity symmetry, which have recognizable incarnations in string theory as end-of-the-world branes and orientifold planes. End-of-the-world branes play a key role in fully resolving the question of why there is no global symmetry generated by the parity Wilson loop.

Finally, we offer some concluding remarks in~\S\ref{sec:conclusions}.

\section{Breaking Discrete Internal Symmetries}
\label{sec:discretegauge}

In this section, we review the physics of domain walls and vortices in a theory with a discrete {\em internal} symmetry, i.e., one that acts locally on fields, and not on spacetime itself. In particular, parity is {\em not} an internal symmetry. We focus on simple examples that illustrate the physics clearly, rather than abstract arguments. First, in ~\S\ref{subsec:exZ2background}, we provide an example of a nontrivial background gauge field for a discrete global symmetry. We then use this example in~\S\ref{subsec:globalDW} to show that when a discrete global symmetry is spontaneously broken, an exactly stable domain wall may be formed. The use of a background gauge field may seem like an unnecessary level of formalism for such a well-known fact, but it allows for a precise analogy in~\S\ref{sec:parityDW} when we turn to the less familiar case of parity symmetry.

Next, we explain the idea of a discrete gauge vortex in~\S\ref{subsec:twistvortex}. This is an object around which fields come back to themselves only up to a gauge transformation. We show that the existence of dynamical vortices, in a theory with a discrete gauge symmetry, renders the domain walls unstable. In a theory of quantum gravity, we expect that any internal discrete symmetry is either explicitly broken or gauged, and in the latter case accompanied by dynamical vortices. Hence, we do not expect to find exactly stable domain walls.

This section contains no new results, but is provided in order to make a clear contrast with the discussion of parity domain walls in the following two sections.

\subsection{Discrete Background Gauge Fields}
\label{subsec:exZ2background}

A theory with a global symmetry $G$ can be coupled to a background gauge field.\footnote{We assume that $G$ does not have an 't~Hooft anomaly.} For discrete global symmetries, this is not a continuously varying field, like the familiar gauge field $A_\mu(x)$ for a continuous symmetry. Instead, the gauge field configuration is specified by choices of group elements $W(C) \in G$ associated to closed loops $C$ in spacetime.\footnote{\label{footnote:nonabelian}To be precise, when $G$ is nonabelian, the group elements $W(C)$ are only defined up to conjugacy. The Wilson loop observables, which are traces in a particular representation $\rho$, are independent of the element in a conjugacy class.} We refer to these group elements as {\em holonomies} or (untraced) {\em Wilson loops}. They are the discrete analogue of Aharonov-Bohm phases: when a particle with $G$ charge circles the loop, it will not return to its original state, but will transform by the group operation $W(C)$. Calling this choice a background {\em gauge} field may be slightly confusing, because we do not have a gauge symmetry. The key word is {\em background}, meaning that the gauge field is a rigid choice that we make once and for all, i.e., a classical source to which the dynamical quantum fields can respond. In other words, the Wilson loops $W(C)$ are not quantum operators, but simply c-numbers. This contrasts with a dynamical gauge field, which we discuss in \S\ref{subsec:twistvortex}.

\begin{figure}[!h]
\centering
\includegraphics [width = 0.4\textwidth]{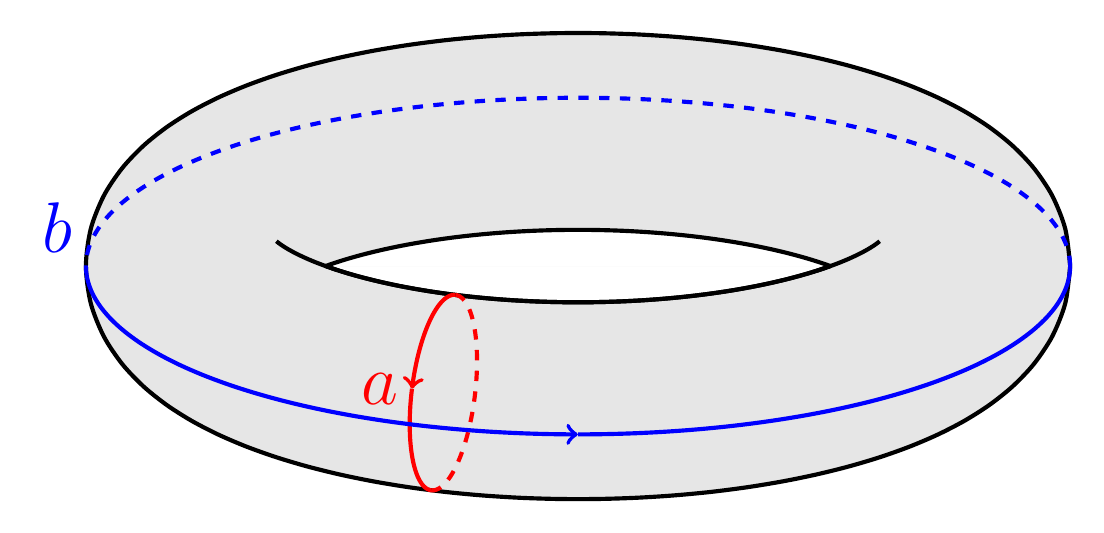}
\caption{
Torus with labeled $a$- and $b$-cycles. In $\ZZ_2$ gauge theory we assign holonomies (Wilson loops) $\pm 1$ on each cycle.
}
\label{fig:torus1}
\end{figure}

For a concrete example, take $G = \ZZ_2$, the group with two elements $\pm 1$, and take space to be a two-dimensional torus $T^2 \cong S_a^1 \times S_b^1$. We refer to the two circles as the $a$-cycle and the $b$-cycle (see Fig.~\ref{fig:torus1}). A background gauge field is a choice of an element of $\ZZ_2$ for each of the two cycles. Consider the case where the background gauge field Wilson loop is $W(a)= -1$ on the $a$-cycle and $W(b) = +1$ on the $b$-cycle.

\begin{figure}[!h]
\centering
\includegraphics [width = 0.45\textwidth]{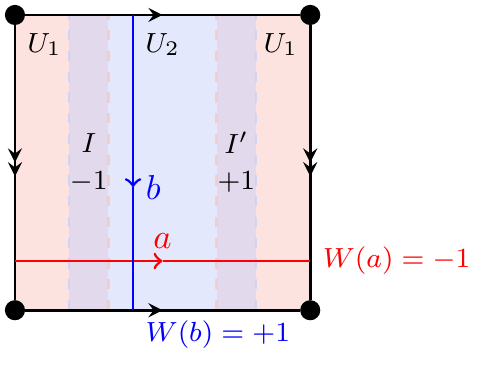}
\caption{
Nontrivial background gauge field configuration on a torus. The left and right edges are identified, as are the top and bottom edges. We show two coordinate patches ($U_1$, orange; $U_2$, purple), with two disjoint overlaps $I$ and $I'$ with transition elements $-1$ and $+1$ respectively. The $a$-cycle passes through both overlap regions, for a nontrivial Wilson loop $W(a) = -1 \times +1 = -1$. The $b$-cycle does not pass through any overlap region, and has trivial Wilson loop $W(b) = +1$.
}
\label{fig:torusgaugefield}
\end{figure}

In such a background gauge field configuration, $\ZZ_2$ charged fields traversing the $a$-cycle pick up an extra minus sign. We often say that such fields are not single-valued. A more careful way to say this is that we cannot describe the field configuration with a single coordinate chart, as in the familiar case of the Dirac monopole. Instead, we could cover our torus by two different coordinate charts, $U_1$ and $U_2$, which overlap in a region $U_{12} = I \sqcup I'$. (The symbol $\sqcup$ means ``disjoint union,'' i.e., $I$ and $I'$ do not intersect.) We show these coordinate charts in Fig.~\ref{fig:torusgaugefield}, on a torus that has been sliced open into a rectangle, glued together along the marked edges. 

Whenever our coordinate charts overlap, we need a {\em gluing rule} that tells us how to relate the coordinates and fields in one chart to those in another chart. For example, a given point $p \in I$ might be given coordinates $x^\mu_1$ in chart $U_1$ and $x^\mu_2$ in chart $U_2$, and we might have a neutral scalar field $\phi$ that is glued together according to the simple rule $\phi_2(x_2) = \phi_1(x_1)$, using notation where $\phi_i$ denotes the representation of the field $\phi$ in the coordinate patch $U_i$.

A background discrete gauge field is a locally constant choice of group element $g_{i \to j} \in G$ assigned to every coordinate chart overlap $U_i \cap U_j$.\footnote{We also require that $g_{j \to k} g_{i\to j} = g_{i \to k}$ on the triple overlaps $U_i \cap U_j \cap U_k$, but this constraint is not relevant in our examples.} In the context of our example, locally constant means that there is one choice of $g_{1 \to 2}$ for the region $I$ and another choice for the region $I'$; moving within a region doesn't change the group element (hence ``constant''), but it can differ on disconnected regions (hence only ``locally'' so). This definition of background gauge fields in terms of choices of group elements on overlaps is related to our earlier definition in terms of assignments of $W(C) \in G$ to closed loops in space as follows. For every loop $C$, one takes $W(C)$ to be the products of the group elements one encounters when traversing the loop $C$. For example, when traversing the $a$-cycle in Fig.~\ref{fig:torusgaugefield}, we encounter two coordinate patch overlaps, and we have $W(a) = -1 \times 1 = -1 \in \ZZ_2$. In contrast, we encounter no transitions when traversing the $b$-cycle, so we have $W(b) = +1 \in \ZZ_2$.

Background gauge field configurations define nontrivial gluing rules for fields transforming in a representation $\rho$ of the group: given the transition element $g_{1 \to 2}$ when moving from chart $U_1$ to chart $U_2$, we have $\phi_2(x_2) = \rho(g_{1 \to 2}) \cdot \phi_1(x_1)$. In the case of a $\ZZ_2$-odd scalar field $\phi$, this simply means that $\phi_2(x_2) = \pm \phi_1(x_1)$ when the overlap region is assigned $\pm 1$. We now show that this gluing rule leads directly to the presence of domain walls in the symmetry-breaking phase.

\subsection{Stability of Global Domain Walls}
\label{subsec:globalDW}

Suppose that the theory we consider spontaneously breaks its $\ZZ_2$ global symmetry: there is a $\ZZ_2$-odd scalar operator $\phi$ with two different ground states $\langle \phi \rangle = \pm v$, exchanged by the $\ZZ_2$ action.\footnote{You could imagine that $\phi$ is an elementary scalar with potential $V(\phi) = \lambda(\phi^2 - v^2)^2$, for concreteness, but $\phi$ could be a composite operator, or the result of nonperturbative dynamics.} We are free to study this theory on a nontrivial background; we choose that of Fig.~\ref{fig:torusgaugefield}.

\begin{figure}[!h]
\centering
\includegraphics [width = 0.45\textwidth]{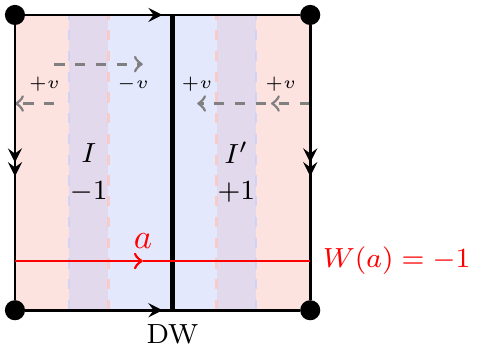}
\caption{
Spontaneous symmetry violating $\ZZ_2$-odd field configuration on the torus with the $\ZZ_2$ Wilson line $W(a) = -1$ introduced in Fig.~\ref{fig:torusgaugefield}. Beginning with the choice $\phi_1 = +v$ on the left, we encounter a clash in the central region between $\phi_2 = -v$ (approaching from the left, with a sign flip in $I$) and $\phi_2 = +v$ (circling around from the right, with no sign flip in $I'$). This indicates the topological necessity of a domain wall wrapping the $b$-cycle, indicated here by the heavy black line DW.
}
\label{fig:torusdomainwall}
\end{figure}

Let's try to write down a vacuum field configuration in our example background gauge field (see Fig.~\ref{fig:torusdomainwall}). At the left edge of the figure, in the patch $U_1$, we set $\phi_1 = +v$. Using the torus topology, we can wrap around the figure from left to right, and so we also have $\phi_1 = +v$ on the right side, still in patch $U_1$. Moving from the right to the middle, we cross the region $I'$ into the chart $U_2$, and the gluing rule tells us that $\phi_2 = +v$. However, moving from the left into the middle, we cross the region $I$ into the chart $U_2$, and the gluing rule tells us that $\phi_2 = -v$. But now we have a clash: we found different answers by taking two different paths around the torus! This clash means that there is no vacuum state with spontaneously broken $\ZZ_2$ symmetry in our chosen background gauge field. Instead, there must be a {\em domain wall}, interpolating between the two vacua $\phi = -v$ and $\phi = +v$. This domain wall wraps around the $b$-cycle, intersecting the $a$-cycle transversely at a single point, as depicted in Fig.~\ref{fig:torusdomainwall} (thick black line).

The lesson is that the presence of a domain wall is a topological requirement for spontaneously breaking a global symmetry in the presence of background gauge fields. Given a nontrivial holonomy $g$ around some loop, we must cross a domain wall between two vacua $v$ and $\rho(g) \cdot v$ when circling the loop. As a consequence, the domain wall must be absolutely stable in the presence of a background gauge field.

Now, we come to a crucial point: locally near the domain wall there is no way to tell if the domain wall is embedded in flat space or is living in a nontrivial background gauge field. We have just argued that, in the nontrivial background, the domain wall is stable: there can be no process that could tear open a hole in the wall. By locality, this holds in {\em every} background! Thus, we have learned something about QFT in flat space by studying QFT in nontrivial backgrounds, which is a useful general technique.

\begin{figure}[!h]
\centering
\includegraphics [width = 0.3\textwidth]{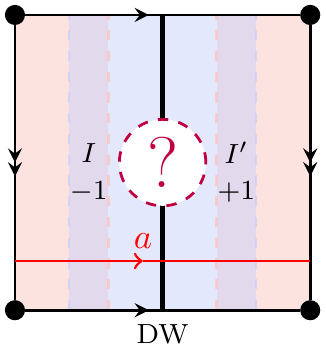}
\caption{
For the domain wall to decay, the gauge field configuration must become dynamical, allowing the transition function to change by some novel process in the region marked with the question mark. We will see in \S\ref{subsec:higgsingdiscrete} that this can actually happen in gauge theories with dynamical vortices.
}
\label{fig:torusdomainwallquestion}
\end{figure}

To escape the conclusion that domain walls are stable, one must either explicitly break the symmetry (lifting the degeneracy of the vacua) or gauge it. Gauging the symmetry means that we sum over gauge fields in the path integral, instead of fixing one as a classical background. As a result, a dynamical process could alter the transition functions in a region that pierces the domain wall, along the lines suggested in Fig.~\ref{fig:torusdomainwallquestion}. To understand how this can happen, we turn our attention to gauge theory.

\subsection{Discrete Gauge Symmetries and Vortices}\label{subsec:twistvortex}

At first glance, discrete gauge symmetries seem difficult to distinguish from discrete global symmetries: they impose the same constraints on terms in the Lagrangian and the same selection rules on correlation functions. There is no propagating gauge boson. However, there is a meaningful distinction between discrete global and gauge symmetries, which will allow for the decay of domain walls associated to the spontaneous breaking of discrete gauge symmetries.

A discrete gauge symmetry allows for the existence of {\em vortices}: dynamical codimension-2 objects in spacetime (strings, in 4d) that have Aharonov-Bohm interactions with particles carrying gauge charge. These can be thought of as the magnetically charged objects of discrete gauge theories. For example, for $\mathbb{Z}_N$ gauge theory, a particle of charge $q \in \mathbb{Z}_N$ circling a vortex of magnetic charge $k \in \mathbb{Z}_N$ acquires a phase $\exp(2\pi \iu q k/N)$~\cite{Krauss:1988zc}. When $\mathbb{Z}_N$ arises by higgsing $U(1)$ with a scalar field of charge $N$, the vortices are solitonic magnetic flux tubes of the $U(1)$ theory with fractional magnetic flux $k/N$, i.e., they are ANO vortices~\cite{Abrikosov:1956sx, Nielsen:1973cs}. 

\begin{figure}[h]
\centering
\includegraphics [width = 0.3\textwidth]{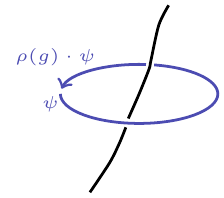}
\caption{
A {\em vortex} is a codimension-2 dynamical object (black curve) that enforces a discrete holonomy $g \in G$, so that a $G$-charged particle $\psi$ circling the vortex (blue path) comes back to itself only up to a gauge transformation. A familiar example is the $\mathbb{Z}_N$ magnetic string~\cite{Krauss:1988zc}.
}
\label{fig:twist_string}
\end{figure}

As a codimension two object, a vortex is encircled by a curve $C$, and enforces a particular gauge holonomy, $W(C) = g \in G$.\footnote{Strictly speaking, the group element $g$ is only defined up to conjugacy, as in Footnote~\ref{footnote:nonabelian}.} As illustrated in Fig.~\ref{fig:twist_string}, this implies that fields $\psi$ transforming in a representation $\rho$ of $G$ are not single-valued. We must adopt multiple coordinate charts with nontrivial transition functions to describe the discrete gauge field surrounding the vortex (cf.~Fig.~\ref{fig:torusgaugefield}), just as we must do to describe the electromagnetic gauge field surrounding a Dirac monopole. 

In a discrete gauge theory, we can always insert a {\em static} vortex along a codimension-2 manifold $\Sigma$. This means that we impose a boundary condition in the path integral, so that we only sum over gauge field configurations with the prescribed holonomy on curves $C$ that link with $\Sigma$. This prescription defines an extended operator of codimension two, which is known as a Gukov-Witten operator~\cite{Gukov:2006jk, Gukov:2008sn}.\footnote{A Gukov-Witten operator is to a dynamical vortex as a Wilson line is to a dynamical charged particle, such as an electron.} More interesting, for our purposes, are {\em dynamical} vortices. If these exist, the path integral includes a sum over all of the codimension-2 worldvolumes they can sweep out in spacetime, together with all of the field configurations with the right boundary conditions on the vortices.

Vortices for internal gauge symmetries can arise for any disconnected group, not just for finite groups. For example, one can study $O(2)$ gauge theory, which arises by gauging the $\mathbb{Z}_2$ charge conjugation outer automorphism symmetry of $U(1)$ gauge theory~\cite{Kiskis:1978ed}. In this case, the vortices are ``Alice strings'': an electron circling such a string comes back as a positron~\cite{Schwarz:1982ec, Preskill:1990bm, Alford:1990mk}. Charge conjugation symmetry is {\em not} analogous to parity or time reversal symmetry, despite their adjacent appearance in QFT textbooks: charge conjugation is an internal symmetry, not a spacetime symmetry. Although Alice strings may seem exotic, in many ways they are quite analogous to familiar ANO strings in superconductors. We will see in \S\ref{sec:parityDWQG} that the situation is quite different for orientation-reversing spacetime symmetries, which do not admit any analogue of a vortex.

In UV complete quantum gravity with a disconnected gauge group $G$, we expect to find a complete spectrum of dynamical vortices, as an application of the general Completeness Hypothesis~\cite{Polchinski:2003bq}. In the absence of dynamical vortices, topological Wilson line operators labeled by representations of $\pi_0(G)$ generate a (possibly non-invertible) $(d-2)$-form symmetry. The absence of global symmetries in quantum gravity then {\em requires} the existence of vortices~\cite{Rudelius:2020orz, Heidenreich:2021xpr}.\footnote{This connects to a very active area of investigation on generalized symmetries in gauge theories. See~\cite{Rudelius:2020orz} for the special case $d = 3$ where vortices are particles,~\cite{McNamara:2021cuo} for the possibility of vortices as gravitational solitons,~\cite{Heidenreich:2021xpr} for a general argument, and~\cite{Casini:2020rgj, Casini:2021zgr, Henning:2021ctv, Roumpedakis:2022aik, Bhardwaj:2022yxj, Arias-Tamargo:2022nlf, Bhardwaj:2022scy, Bartsch:2022mpm} for other recent work involving completeness and/or outer automorphism symmetries, their gauging, and the associated non-invertible global symmetries.} 

\subsection{Vortices and Instability of Domain Walls}
\label{subsec:higgsingdiscrete}

\begin{figure}[h]
\centering
\includegraphics [width = 0.4\textwidth]{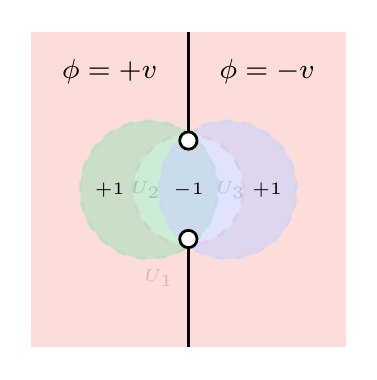}~~~\includegraphics [width = 0.35\textwidth]{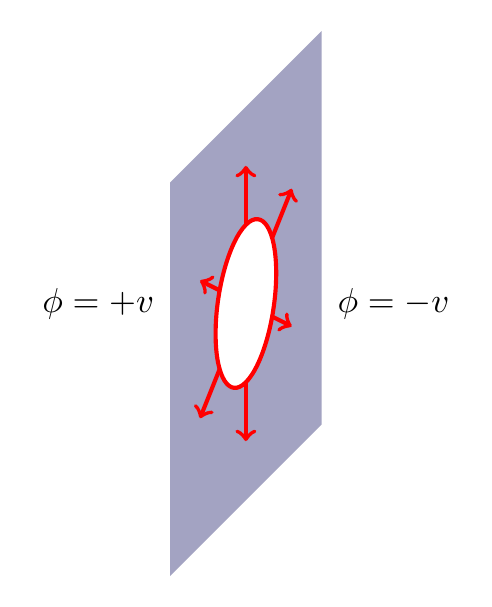}
\caption{
Domain wall and vortex for (internal) $\mathbb{Z}_2$ gauge theory. {\bf Left:} the domain wall is the vertical black line, ending on white punctures corresponding to a vortex piercing the plane of the page. Three coordinate charts $U_i$ are indicated by shaded regions. On the overlaps, we have chosen transition functions labeled by elements of $\mathbb{Z}_2$: $+1$ on $U_1 \cap U_2$ and $U_1 \cap U_3$, and $-1$ on $U_2 \cap U_3$. The product, $-1$, is an invariant corresponding to the holonomy around the vortex. It is this nontrivial holonomy that allows the domain wall to end on the vortex. {\bf Right:} a visualization of the domain wall (shaded blue plane) being destroyed by nucleation of a vortex (red circle), which can cut an expanding hole in the domain wall. The left panel is a slice through the right panel.
}
\label{fig:Z2DW}
\end{figure}

\begin{figure}[!h]
\centering
\includegraphics [width = 0.95\textwidth]{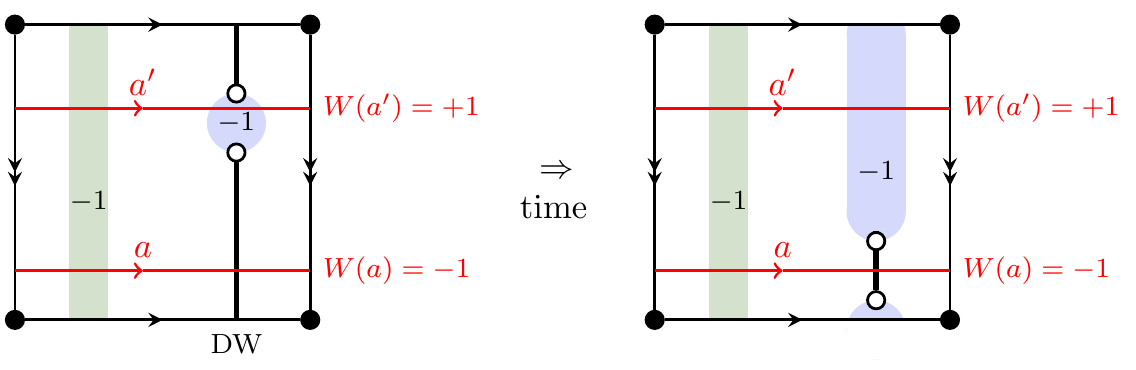}
\caption{
The ability to alter gauge transition functions using a vortex circumvents the topological argument for domain wall stability. For clarity, we do not show the full set of coordinate charts, only the overlap regions where the transition functions are nontrivial (green and purple shaded areas). {\bf Left:} shortly after the hole in the domain wall nucleates, most horizontal Wilson loops are still $-1$; only a small region where the Wilson loop threads the gap in the domain wall has trivial holonomy, as indicated along the cycle $a'$. {\bf Right:}  The vortex expands outward, cutting open larger and larger regions in which the holonomy around the horizontal cycle is now $+1$ rather than $-1$. After the vortex annihilates away, the gauge field configuration will be entirely trivial.
}
\label{fig:torusdomainwalldecay}
\end{figure}

A global symmetry connects {\em different} physical states, whereas a gauge symmetry does not: the vacuum on opposite sides of a domain wall is in fact the {\em same} physical vacuum. This allows for the possibility of physical processes that cut a hole in a domain wall and connect the physically equivalent states on the two sides. This process is enabled by vortices: a domain wall can end on a vortex~\cite{Kibble:1982ae, Vilenkin:1982ks, Kibble:1982dd, Everett:1982nm}, and conversely, every vortex in the symmetry-breaking phase must be attached to a domain wall. This is illustrated in the left panel of Fig.~\ref{fig:Z2DW} for the case of $\mathbb{Z}_2$ gauge theory. A vortex inserts a nontrivial holonomy, which allows us to interpolate between the field configurations $\phi = +v$ and $\phi = -v$ by passing through a coordinate patch overlap on which $\phi$ in one patch is identified with $-\phi$ in the other, {\em without} passing through a domain wall. The left panel of Fig.~\ref{fig:Z2DW} can be embedded in the ``?'' in Fig.~\ref{fig:torusdomainwallquestion}. We depict this in Fig.~\ref{fig:torusdomainwalldecay}, showing that nucleating a vortex opens a band in the torus on which the holonomy around the horizontal cycle becomes $W(a') = -1 \times -1 = +1$. This region can expand (if it is larger than a critical radius), until the nontrivial holonomy is erased.

Spontaneous breaking of a discrete gauge symmetry leads to the formation of domain walls in cosmology~\cite{Zeldovich:1974uw, Kibble:1976sj, Zurek:1985qw}. The endability of domain walls on vortices, which are strings in 4d, allows for two different cosmological mechanisms by which a network of domain walls can be destroyed after a discrete gauge symmetry violating phase transition~\cite{Kibble:1982ae}. The first is if the universe was already filled with a network of dynamical strings before the phase transition occurs. (For example, if the discrete gauge group arose from higgsing a larger continuous gauge group, a network of strings would have been created in the earlier higgsing phase transition through the Kibble-Zurek mechanism.) The discrete symmetry-breaking transition creates a network of domain walls that end on the strings, and which exert a confining force on the strings, drawing them together until they annihilate and the whole string-wall network is destroyed~\cite{Vilenkin:1982ks}. The second is if the domain walls form in a universe without a population of strings. In this case, it is still possible for quantum tunneling (essentially, the Schwinger process~\cite{Schwinger:1951nm}) to nucleate a loop of string inside a domain wall, bounding a hole in the wall, as illustrated in the right panel of Fig.~\ref{fig:Z2DW}. This loop of string can then expand (if this is energetically favorable) and eat up the wall. Whether this process is fast enough, in practice, to avoid a cosmological problem depends on the relative size of the string tension $\mu$ and the domain wall tension $\sigma$. The rate is exponentially slow in the dimensionless ratio $\mu^3/\sigma^2$, so it will be negligible if these scales are well-separated~\cite{Kibble:1982dd}. 

Given the above discussion, we expect that domain walls formed by the spontaneous breaking of internal discrete symmetries are always unstable in quantum gravity. Because quantum gravity has no global symmetries, the discrete symmetry must either be broken explicitly (lifting the degeneracy of the vacua and causing the domain walls to collapse) or gauged (rendering the domain walls unstable due to vortices). However, we will see below that this argument is limited to internal symmetries: domain walls formed from the breaking of orientation-reversing spacetime symmetries are absolutely stable.

\section{Breaking Parity Symmetry}
\label{sec:parityDW}

In \S\ref{sec:discretegauge}, we saw that domain walls arising from the spontaneous breaking of internal global symmetries are stable, while those arising from gauge symmetries are unstable. In this section, we investigate the analogous question for the spontaneous breaking of parity. Our arguments will parallel those of the previous section as precisely as possible. We begin by discussing non-orientable manifolds as the analogue of nontrivial background gauge fields in~\S\ref{subsec:kleinintro}. We then consider breaking parity on a non-orientable manifold in~\S\ref{subsec:globalparityDW}, leading to a proof of the stability of parity domain walls when parity is a global symmetry, i.e., when we study a parity-symmetric theory on a fixed spacetime. In~\S\ref{subsec:walls_w1}, we describe the relationship between parity domain walls and the first Stiefel-Whitney class, which quantifies the necessity of parity domain walls in a fixed, non-orientable spacetime.

As a result, any process by which a parity domain wall could decay must involve a change in the spacetime topology. This conclusion is directly analogous to the observation in~\S\ref{subsec:globalDW} that any instability of a domain wall for an internal discrete gauge symmetry would require a change in the gauge field configuration. For internal symmetries, the key object in the theory allowing such a transition was the dynamical vortex introduced in \S\ref{subsec:twistvortex}. This is where the story of parity diverges from the internal symmetry case. In fact, there is no analogous process, and parity domain walls are stable in quantum gravity, even after allowing for topology change. We explain how to properly frame this question in~\S\ref{sec:parityDWQG} in terms of boundary conditions for the gravitational path integral.

\subsection{Background Parity Gauge Fields and the Klein Bottle}
\label{subsec:kleinintro}

The logic of this section will parallel that of \S\ref{sec:discretegauge}, where we began in \S\ref{subsec:exZ2background} by introducing an example of a nontrivial background $\ZZ_2$ gauge field for an internal global symmetry. For global parity symmetry, the analogue is to define the theory in the presence of a fixed, background {\em parity gauge field}. In more familiar language, this means placing the theory on a non-orientable spacetime manifold. 

A general, unoriented spacetime manifold can be covered by a set of coordinate charts, each with its own local orientation. Parity can be thought of as a specific $\ZZ_2$ gauge field on such a manifold. In  \S\ref{subsec:exZ2background}, we defined a $\ZZ_2$ background gauge field by choosing a locally constant element of $\ZZ_2$ on every coordinate chart overlap. The special case of parity is one where {\em we cannot freely choose the $\ZZ_2$ element, but fix it based on the coordinate charts}. Specifically, we define the {\em parity  gauge field} $g^{\rm P}$ on an overlap $U_i \cap U_j$ by the transition functions
\begin{equation}\label{w1_def}
g^{\rm P}_{i \to j} = \mathrm{sign}\, \det \left(\frac{\partial x^\nu_j}{\partial x^\mu_i}\right) \in \ZZ_2,
\end{equation}
the sign of the Jacobian determinant for the coordinate transformation on the overlap of $U_i$ and $U_j$. In other words, we pick $g^{\rm P}_{i \to j} = +1$ on a given overlap region if the coordinate charts have the same orientation in that region, and $g^{\rm P}_{i \to j} = -1$ if they have the opposite orientation. The parity gauge field $g^{\rm P}$ may be viewed as a composite $\ZZ_2$ gauge field, built out of the gluing data of the spacetime manifold itself.

Just as we did for ordinary background $\ZZ_2$ gauge fields in \S\ref{subsec:exZ2background}, we can assign to any closed loop $C$ in our spacetime manifold a $\ZZ_2$ element, which we can think of as a parity Wilson loop $W_{\rm P}(C)$. This element is given by the product of all of the $g^{\rm P}_{i \to j} = \pm 1$ transition elements we encounter as we circle around the path. For an {\em orientable} manifold, every loop is assigned the $\ZZ_2$ element $+1$: we encounter no net orientation reversal around any cycle. In that case, we can always flip our orientation choice on some coordinate charts to make the orientation the same everywhere. On the other hand, a {\em non-orientable} manifold is one in which there is a loop assigned to the $\ZZ_2$ element $-1$. In that case, no amount of redefining our coordinate charts can ever eliminate this global orientation reversal. Just like the Wilson loops for a background internal gauge field, the parity Wilson loop $W_{\rm P}(C)$ is simply a c-number when parity is a global symmetry and the topology of spacetime is fixed.

\begin{figure}[h]
\centering
\includegraphics [width = 0.4\textwidth]{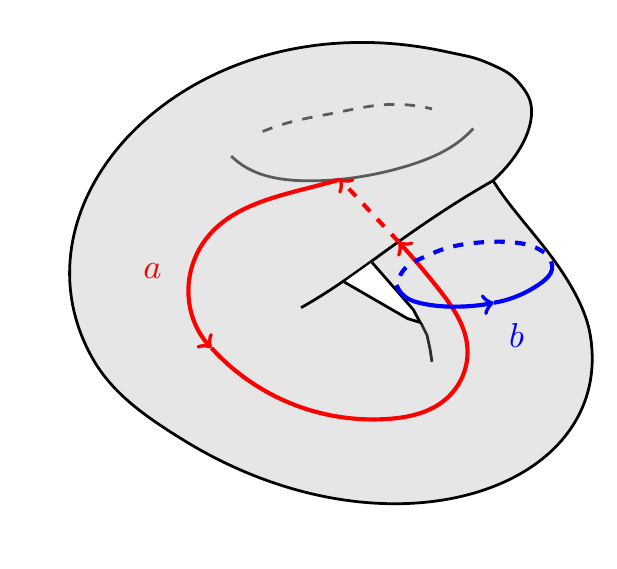}\quad\quad\includegraphics [width = 0.49\textwidth]{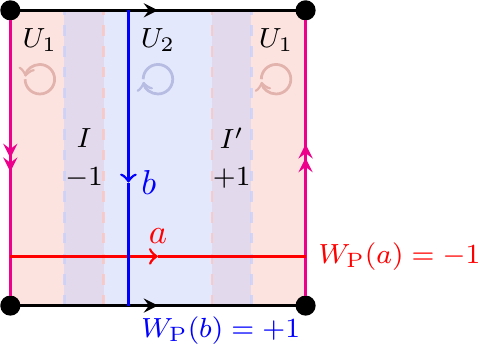}
\caption{
The Klein bottle. {\bf Left:} immersion of the Klein bottle in 3d space, with the orientation-reversing $a$-cycle indicated in red and the orientation-preserving $b$-cycle in blue. 
{\bf Right:} coordinate charts on a Klein bottle. This figure is largely analogous to the torus depicted in Fig.~\ref{fig:torusgaugefield}, with the crucial difference that the left and right edges (highlighted in magenta) are now identified with opposite orientation. We assign each coordinate patch an orientation, indicated by the clockwise or counterclockwise circle shown near the upper edge of the square. Importantly, the parity transition functions ($-1$ on $I$ and $+1$ on $I'$) cannot be freely chosen, but are entirely determined by the orientations of the overlapping coordinate charts. When traversing the $a$-cycle, we encounter a net orientation reversal, corresponding to the parity Wilson line $W_{\rm P}(a) = -1$. 
}
\label{fig:klein}
\end{figure}

Let us illustrate the ideas discussed above in a concrete example, taking space to be the Klein bottle $K$. We show an explicit example of coordinate charts on $K$ in Fig.~\ref{fig:klein}. Notice that the diagram of $K$ as a square with glued edges is very similar to that for the torus, but the left and right edges are now glued together with a relative flip. As we did for the torus in \S\ref{subsec:exZ2background}, we choose two coordinate patches $U_1$ and $U_2$ that intersect in two disconnected regions, $U_1 \cap U_2 = I \sqcup I'$. We assign a local orientation on each patch; for the choice made in the figure, the transition function in $I$ is orientation-reversing, whereas that in $I'$ is orientation-preserving. The parity gauge field \eqref{w1_def} is given by
\begin{equation}\label{uP_klein}
\left. g^{\rm P}_{1\to2} \right|_{I} = -1, \quad \left. g^{\rm P}_{1\to2} \right|_{I'} = +1.
\end{equation}
Around the cycle $a$ of the Klein bottle (red curve in Fig.~\ref{fig:klein}), we have a parity Wilson loop $W_\mathrm{P}(a) = \left. g^{\rm P}_{1\to2} \right|_{I} \left. g^{\rm P}_{2\to1} \right|_{I'} = -1$, so there is a net orientation change around this cycle, establishing the familiar fact that $K$ is non-orientable.

Just as we discussed for ordinary discrete symmetries in \S\ref{subsec:exZ2background}, the choices of $\ZZ_2$ group elements $g^{\rm P}_{i \to j}$ on overlaps play a role in defining gluing rules for fields on our space. A parity-odd scalar field, also known as a {\em pseudoscalar} field, is one that behaves as $\phi_j(x_j) = g^{\rm P}_{i \to j} \cdot \phi_i(x_i)$ on an overlap $U_i \cap U_j$. In other words, if Alice chooses a local right-handed coordinate system and Bob chooses a local left-handed coordinate system in an overlapping region, they will disagree about the sign of a pseudoscalar field.

For example, consider a pseudoscalar field $\phi$ defined on the Klein bottle. The choice of local orientations shown in Fig.~\ref{fig:klein} implies that $\phi_2 = -\phi_1$ in the region $I$. We could have made an alternative choice, e.g., taking the coordinate system on $U_2$ to have the opposite orientation, which would have made the transition function on $I'$ orientation-reversing but that on $I$ orientation-preserving. The invariant fact, independent of our choice of coordinate patches, is that because $W_\mathrm{P}(a) = -1$, $\phi$ will encounter a net sign change when circling $a$.

Gluing rules for parity transition functions can be extended to other kinds of fields, e.g., a left-handed fermion field in one patch should glue to a right-handed one in an overlapping patch of opposite orientation (perhaps with charges inverted, as in CP in the Standard Model). Defining fermions on non-orientable manifolds requires a choice of either pin$^+$ or pin$^-$ structure, but the details will not be relevant to our main arguments in this paper.

\subsection{Breaking Global Parity on a Klein bottle}
\label{subsec:globalparityDW}

Our next step is the direct analogue of \S\ref{subsec:globalDW}, where we used a nontrivial $\ZZ_2$ background gauge field to argue that the domain wall for a spontaneously broken global $\ZZ_2$ symmetry is stable. Here, we will argue that a domain wall for global parity symmetry is stable by studying spontaneous parity breaking on the Klein bottle. 
We suppose that our QFT contains a pseudoscalar operator $\phi$, with vacua at $\langle \phi \rangle = \pm v$, spontaneously breaking parity.\footnote{As in \S\ref{subsec:globalDW}, one could imagine that $\phi$ is an elementary pseudoscalar with a potential $\lambda(\phi^2 - v^2)^2$ for concreteness. However, this example is awkward, as it has a {\em separate} internal $\ZZ_2$ symmetry $\phi(x) \mapsto -\phi(x)$ (acting trivially on other fields), which is not parity. One should add some other terms to break that internal $\ZZ_2$ symmetry and preserve parity, such as a parity-odd Yukawa coupling to fermions, to avoid confusing matters.}

\begin{figure}[h]
\centering
\includegraphics [width = 0.3\textwidth]{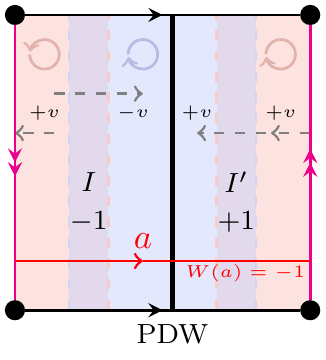}\quad\quad\quad\includegraphics [width = 0.3\textwidth]{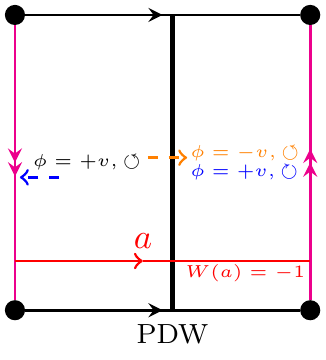}
\caption{
Parity domain wall configuration (thick solid black line) on a Klein bottle. 
{\bf Left:} The coordinate charts are as previously depicted in Fig.~\ref{fig:klein}. 
As in the earlier case of internal symmetries in Fig.~\ref{fig:torusdomainwall}, the domain wall and the cycle $a$ with a parity Wilson loop (red line) intersect transversely at a single point. 
{\bf Right:} a different view, without reference to explicit coordinate charts but showing the gauge-equivalent configurations $(+v, \circlearrowright)$ and $(-v, \circlearrowleft)$ as viewed by Alice (blue path) and Bob (orange path) respectively, upon reaching the right-hand side given the same starting point on the left. Bob passes through the domain wall, but Alice does not.
}
\label{fig:kleinDW}
\end{figure}

Now, we can search for parity-violating configurations of the pseudoscalar $\phi$ on the manifold $K$. We show an example in the left panel of Fig.~\ref{fig:kleinDW}, making use of the coordinate charts previously illustrated in Fig.~\ref{fig:klein}. We could start, for example, near the left-hand boundary in patch $U_1$ with the choice $\phi_1 = +v$. Moving to the left and passing through the glued magenta boundary, this remains $\phi_1 = +v$ near the right-hand boundary, and after moving further left through the region $I'$ with positive transition function into the region $U_2$ becomes $\phi_2 = +v$. On the other hand, moving to the right, we encounter an orientation-reversing transition in the region $I$ and thus have $\phi_2 = -v$. The only way to make sense of this result is with a domain wall between $I$ and $I'$.  When crossing through it, the value of $\phi_2$ interpolates from $+v$ to $-v$. The domain wall, illustrated with a thick black vertical line in Fig.~\ref{fig:kleinDW}, intersects the nontrivial parity Wilson loop (red line) transversely, just as we saw for internal symmetries in \S\ref{subsec:globalDW}.

We show a different (and possibly more intuitive) way of thinking about the same result in the right panel of Fig.~\ref{fig:kleinDW}. Here, without discussing explicit coordinate patches, we consider parity as the identification of (pseudoscalar field value, orientation) pairs: ${(\phi, \circlearrowleft)} \sim {(-\phi, \circlearrowright)}$. Suppose that Alice and Bob agree on the orientation $\circlearrowleft$ and the field value $\phi = +v$ at some point near the left edge of the square depicted in the right panel of Fig.~\ref{fig:kleinDW}. Alice follows the dashed blue path off the left-hand edge and wraps around to the right. She sees the field value to be constant along this path, but because she crosses the orientation-reversed glued edge, at the meeting point she sees the pair $(+v, \circlearrowright)$. Bob takes the dashed orange path directly across the middle of the Klein bottle, so his orientation remains $\circlearrowleft$. He can only agree with Alice if his field value is now $-v$: that is, he sees the field and orientation as $(-v, \circlearrowleft)$ because he has crossed through a domain wall on his way to meet Alice.

The existence of the domain wall on the Klein bottle is a topological constraint. It is completely analogous to the need for a domain wall in the nontrivial $\ZZ_2$ background gauge field on a torus that we discussed in \S\ref{subsec:globalDW}. Any manifold on which we can define a nowhere vanishing pseudoscalar field configuration is orientable. Indeed, such a field configuration defines an orientation on the manifold! On a non-orientable manifold, any pseudoscalar field {\em must} vanish along some locus in the manifold (we identify this topological class of this locus in~\S\ref{subsec:walls_w1}). Given a parity-violating potential $V(\phi)$, the vanishing locus of $\phi$ is a domain wall.  

This argument establishes that no local process in quantum field theory could destroy the domain wall. In other words, it establishes that domain walls are stable when defined on a {\em fixed} spacetime manifold. This is the case of parity as a global symmetry. For an ordinary (internal) global symmetry, we saw in \S\ref{subsec:globalDW} that domain walls are stable in the  presence of a fixed background gauge field. On the other hand, gauging the symmetry, and thus allowing the gauge field configuration to change, enabled a decay process for the domain wall, as we saw in \S\ref{subsec:higgsingdiscrete}. The analogue, for parity, would be a process that changes the parity gauge field configuration; however, because it is defined, as in~\eqref{w1_def}, in terms of the choice of spacetime, this means that {\em any process that can destroy a parity domain wall must involve a change in the underlying spacetime manifold}. Such a process necessarily requires quantum gravity, not just quantum field theory. This already leads us to expect that any such process would be extremely suppressed, as we do not encounter macroscopic changes in spacetime topology in the world around us, despite the existence of dynamical gravity. However, we will see in~\S\ref{sec:parityDWQG} that a stronger statement is true: there is {\em no} analogue, for parity domain walls, of the vortex-induced decay process described in \S\ref{subsec:higgsingdiscrete}, even if we allow for dynamical topology change.

\subsection{Parity Domain Walls and the First Stiefel-Whitney Class}\label{subsec:walls_w1}

The argument that we presented for the Klein bottle carries over to any non-orientable, $d$-dimensional spacetime manifold $X$. By definition, it is impossible to define a nowhere-vanishing pseudoscalar field configuration on such a manifold, and so there must be a domain wall along some codimension-1 locus $D$ in spacetime. However, we can say more: we can identify the homology class $[D] \in H_{d-1}(X; \ZZ_2)$ of the domain wall in terms of a topological invariant of the spacetime manifold $X$.

This topological invariant is the first Stiefel-Whitney class of the tangent bundle of $X$, which  is denoted $w_1(TX) \in H^1(X; \ZZ_2)$. Often one simply writes $w_1(X)$, leaving implicit the tangent bundle, or even $w_1$ if the spacetime manifold $X$ is clear. One way to define the first Stiefel-Whitney class is via the parity gauge field $g^{\rm P}$, introduced above in \S\ref{subsec:kleinintro}. In general, gauge-equivalence classes of $G$ gauge fields on a manifold $X$ are in one-to-one correspondence with elements of the cohomology $H^1(X, G)$. Thus, the parity gauge field $g^{\rm P}$ defined in \eqref{w1_def} corresponds to a cohomology class $[g^{\rm P}] \in H^1(X; \ZZ_2)$, which is precisely the first Stiefel-Whitney class of $X$. The first Stiefel-Whitney class is the topological obstruction to orientability. If $w_1 = 0$, then one may define an oriented atlas on $X$ by flipping the orientations of some coordinate patches, trivializing the parity gauge field $g^{\rm P}$. In contrast, if $w_1 \neq 0$, then no such oriented atlas exists, $g^{\rm P}$ is a nontrivial gauge field, and $X$ is non-orientable.

Associated to the cohomology class $w_1$ is a Poincar\'e dual homology class $[D] \in H_{d-1}(X;\ZZ_2)$ of codimension one in spacetime. Any pseudoscalar field configuration vanishes along a codimension-1 domain wall $D$ that is a representative of this homology class. In terms of this homology class, we may represent the parity Wilson loop $W_{\rm P}(C)$ as follows. Given a closed curve $C$ in spacetime, choose a representative $D$ of $[D]$ which intersects $C$ transversely. Then traversing $C$ is orientation-preserving if $C$ intersects $D$ an even number of times, while it is orientation-reversing if $C$ intersects $D$ an odd number of times. That is to say
\begin{equation}\label{def_parity_Wilson}
W_{\rm P}(C) = (-1)^{I(C,D)} = (-1)^{\int_C w_1(TX)},
\end{equation}
where $I(C,D)$ is the intersection number, and $\int_C w_1(TX) \in \{0, 1\}$ is the natural pairing between curves and $\ZZ_2$-cohomology classes. In other words, the parity Wilson loop on a curve $C$ simply counts the number of parity domain walls (modulo two) one would need to cross when traversing $C$ in a parity-breaking phase.

\subsection{Stability of Parity Domain Walls in Quantum Gravity}
\label{sec:parityDWQG}

In \S\ref{subsec:globalparityDW}, we argued that parity domain walls are exactly stable in quantum field theory, and could only potentially become unstable due to topology-changing processes in quantum gravity. In particular, we might expect that parity domain walls could be rendered unstable by ``parity vortices'': codimension-2 defects in spacetime, such that traveling around such a defect would bring you back with a parity flip \cite{Choi:1992xp}.\footnote{In \cite{Choi:1992xp}, such objects were called ``CP strings.''} In \S\ref{subsec:twistvortex}, we saw that vortices for an internal, discrete gauge symmetry are associated with a codimension-2 boundary condition for the path integral over discrete gauge fields. Analogously, a parity vortex in a gravitational theory with gauged parity would correspond to a codimension two boundary condition for the gravitational path integral. While it is unclear exactly how to sum over topologies in a UV complete theory of quantum gravity, for our purposes it suffices to consider the semiclassical gravitational path integral.

\begin{figure}[h]
\centering
\includegraphics [width = 0.2\textwidth]{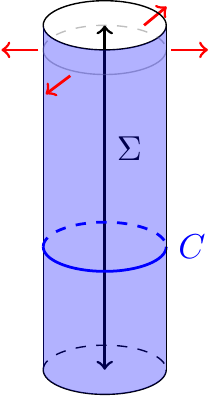}
\caption{
Illustration of a codimension-2 boundary condition~\eqref{eq:BC} on the quantum gravity path integral, representing a static vortex. The vortex worldvolume lies along a $(d-2)$-manifold $\Sigma$ (black line). The boundary of spacetime runs along $C \times \Sigma$ (blue surface), and the gravitational path integral sums over all spacetimes and fields on the exterior (red arrows) obeying the boundary condition.
}
\label{fig:BC}
\end{figure}

Thus, let us consider a gravitational theory in spacetime dimension $d$ with gauged parity, in which we sum over both orientable and non-orientable manifolds. A codimension two boundary condition is a prescription that instructs us to sum not over closed $d$-manifolds, but instead $d$-manifolds $X$ with a prescribed boundary of the form
\begin{equation} \label{eq:BC}
\partial X = C \times \Sigma,
\end{equation}
where $\Sigma$ is the codimension-2 worldvolume of our defect, and $C$ is an infinitesimally small closed curve, or 1-manifold, that defines our boundary condition (see Fig.~\ref{fig:BC}).\footnote{\label{footnote:nontrivial_fibering}More generally, one could consider a nontrivial fibration of $C$ over $\Sigma$, corresponding to a defect on a codimension-2 locus $\Sigma$ with nontrivial normal bundle (or normal cone, if our defect is a spacetime singularity). This simplification will not affect our conclusions. Crucially, what we cannot do is consider a nontrivial fibration of $\Sigma$ over $C$, as such a boundary condition will not correspond to a defect of codimension two.} A parity vortex would correspond to a choice of closed $1$-manifold $C$ such that going around $C$ would bring us back with the opposite orientation. In the presence of a parity-breaking field configuration, a parity vortex would serve as the end of a single parity domain wall. If our gravitational theory had parity vortices as dynamical defects, they would serve to destabilize parity domain walls, just as in \S\ref{subsec:higgsingdiscrete}.

The punchline is that there is no such closed $1$-manifold, as every closed $1$-manifold is orientable! Indeed, every closed 1-manifold is a disjoint union of circles, and the circle is orientable. As a result, there is no boundary condition for the gravitational path integral that could possibly correspond to a parity vortex. The upshot of this simple topological observation is that parity vortices are a mathematical contradiction. They cannot exist, and parity domain walls are exactly stable even once dynamical topology changing processes are allowed.

We may alternatively phrase this result as the fact that every closed $1$-manifold $C$ has a trivial first Stiefel-Whitney class $w_1(TC)$. As discussed in \S\ref{subsec:walls_w1}, parity domain walls on a fixed manifold live on a codimension one homology class $[D]$ dual to $w_1$. As $w_1$ of any closed $1$-manifold must vanish, the number of parity domain walls ending on any codimension-2 defect must be even. In other words, the number of parity domain walls is conserved modulo two, and there is an exact $(d-2)$-form $\ZZ_2$ ``parity domain wall number'' symmetry, which cannot be broken by any codimension-2 defect. We denote this symmetry by $\ZZ_2^{\rm PDW}$, for ``parity domain wall.'' We have met the associated symmetry defect operator~\cite{Gaiotto:2014kfa} before: it is precisely the parity Wilson loop $W_{\rm P}(C)$, a topological line operator that counts the number of domain walls (modulo two) transverse to a closed curve $C$. Now that we have gauged parity, and summed over different spacetime topologies in the path integral, the parity Wilson loop $W_{\rm P}(C)$ is an actual quantum operator. We discuss this symmetry further in~\S\ref{sec:gauge_charge}, and argue that despite the existence of the nontrivial topological line operator $W_{\rm P}$, the symmetry $\ZZ_2^{\rm PDW}$ is nevertheless gauged.

\section{Implications for Model Building}
\label{sec:models}

\begin{figure}[h]
\centering
\includegraphics [width = 0.8\textwidth]{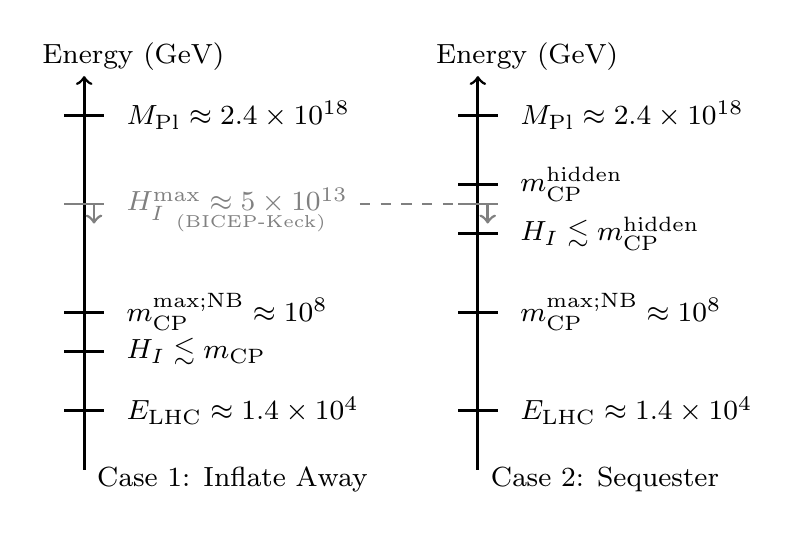}
\caption{
Energy scales in two cosmological scenarios for spontaneous CP violation in Nelson--Barr models, where the maximum scale of CP violation $m_\mathrm{CP}$ is about $10^8\,\mathrm{GeV}$. We bracket the range of energies with two reference scales, the Planck scale $M_\mathrm{Pl}$ on the high end and the LHC center-of-mass energy  for proton collisions on the low end. The empirical upper bound on the inflationary Hubble scale $H_I$ from BICEP-Keck observations is shown in grey. {\bf Left:} Case 1 shows the scenario where the domain walls produced by Nelson--Barr spontaneous CP violation must be inflated away, requiring a low inflationary Hubble scale $H_I$, far below the empirical bound. In this case a large amount of physics (CP violation, inflation, reheating, baryogenesis) must lie in the few decades of energy above current experiments. {\bf Right:} Case 2 considers spontaneous CP violation in a sequestered hidden sector. This can happen at a very high energy $m^\mathrm{hidden}_\mathrm{CP}$ and the domain walls can be inflated away by a large $H_I$. Then a secondary spontaneous CP violation at lower energies produces unstable domain walls. Both scenarios have the potential to predict interesting observables, including gravitational wave signals of decaying domain walls in Case 2.
}
\label{fig:scales}
\end{figure}

In a theory with spontaneously broken parity, domain walls will be formed during a parity-violating phase transition in the early universe. Because these domain walls are absolutely stable, and redshift more slowly than matter or radiation, they pose a cosmological disaster~\cite{Zeldovich:1974uw, Kibble:1976sj} unless they are inflated away~\cite{Guth:1980zm}. If one wishes to postulate an exact CP symmetry in the early universe, for instance to solve the Strong CP problem, this suggests that the scale of inflation (and the reheating temperature) should be below the scale of spontaneous CP violation in order to alleviate the domain wall problem. Depending on the details of the model of spontaneous CP violation, this could be a stringent constraint on inflation. For example, the most recent constraint on primordial tensor modes from the BICEP-Keck collaboration, $r < 0.036$ at 95\% confidence~\cite{BICEP:2021xfz}, translates (in single field slow roll inflation models) to an upper bound on the inflationary Hubble scale of $H_I \lesssim 5 \times 10^{13}\,\mathrm{GeV}$. By contrast, requiring that the value of $\bar \theta$ induced by dimension-five Planck-suppressed operators in typical Nelson-Barr models not spoil the solution of the Strong CP problem translates into a low CP violation scale: $m_\mathrm{CP} \lesssim 10^8\,\mathrm{GeV}$~\cite{Choi:1992xp,Dine:2015jga}. The requirement, in such a model, that $H_I \lesssim m_\mathrm{CP}$ to inflate away domain walls is then nearly six orders of magnitude stronger than the direct experimental bound on $H_I$! Furthermore, it brings the scale of inflation to within a few orders of magnitude of the energies probed by precision flavor and CP experiments. This clearly requires a reassessment of such models, focusing on how to accommodate spontaneous CP violation, inflation, and baryogenesis all within just a few decades of energy. We illustrate the characteristic energy scales of this scenario in the left panel of Fig.~\ref{fig:scales}.

Is there any alternative? Domain walls could also be rendered unstable if there is {\em explicit} CP violation. Then the different vacua will have slightly different energy, and the higher-energy vacuum exerts a force that can cause the domain walls to collapse in on themselves, destroying the network. It has been suggested that parity could be an approximate symmetry, explicitly broken only through higher dimension, Planck-suppressed operators~\cite{Craig:2020bnv}. However, such a scenario is not plausible without some underlying symmetry explanation. If CP or parity is simply not a symmetry at all, there is no reason for it to be broken only through higher dimension operators; one would expect it to be significantly violated at the renormalizable level. One should not think of a theory of quantum gravity as a QFT, which can have exact symmetries, coupled to gravity, which breaks them slightly. Instead, the full theory simply has no symmetries except gauge symmetries. 

Good approximate symmetries in the IR can emerge from other gauge symmetries in the UV. For instance, approximate $B+L$ symmetry in the Standard Model arises not because $B+L$ itself is gauged (it cannot be, because of an ABJ anomaly), but because there are no $(B+L)$-violating renormalizable operators that are invariant under the Standard Model gauge group. Approximate symmetries can also emerge from locality in extra dimensions, which suppresses couplings among fields localized in different places. Finally, they may be enforced by accidental cancellations; e.g., the smaller the Higgs VEV is tuned to be, the lighter the fermions and the better the approximate chiral symmetries below the electroweak scale. None of these mechanisms assumes an approximate global symmetry in the UV, and in fact, any global symmetries are expected to be {\em badly} broken at the UV cutoff scale~\cite{Nomura:2019qps, Cordova:2022rer}. We are unaware of any mechanism for emergent, approximate low-energy CP symmetry, short of fine-tuning each CP-violating coupling individually, that does not begin with an exact (gauged) UV orientation-reversing spacetime symmetry. From this viewpoint, simply postulating Planck-suppressed explicit CP breaking is a non-starter (see related remarks in~\cite{Draper:2022pvk}).

There is, nonetheless, a possible principled route toward a scenario that {\em effectively} has small explicit CP violation, which could be the starting point for further model building. Consider a theory that has an exact CP symmetry, and in which there is a hidden sector of particles that interact extremely weakly with Standard Model fields; for instance, they could be effectively sequestered by separation in extra dimensions. (For sequestering in other contexts, and challenges in realizing it, see, e.g.,~\cite{Randall:1998uk, Luty:2001zv, Schmaltz:2006qs, Kachru:2007xp, Berg:2010ha}.)  Now, the hidden sector could spontaneously break CP symmetry, producing exactly stable domain walls. This must happen at high energies, so that the domain walls could be inflated away. At lower energies, the weak interactions of the hidden sector with the visible sector would manifest as {\em apparent} small explicit CP violation in the visible sector. There is no reason for this to take precisely the form of Planck-suppressed operators, as in~\cite{Craig:2020bnv}, but it could have similar effects. It could more closely resemble explicit {\em soft} breaking of parity or CP, as studied in~\cite{deVries:2021pzl}, for the same reason that soft SUSY breaking is often taken as a model for the visible effects of hidden sector spontaneous SUSY breaking. Visible sector dynamics then could spontaneously break CP a second time, producing a new network of domain walls that would collapse due to the effective explicit CP violation. It is this second spontaneous CP breaking to which the model-dependent bounds on $m_\mathrm{CP}$ would apply. The characteristic energy scales of this scenario are illustrated in the right panel of Fig.~\ref{fig:scales}.

This scenario has the potential to allow the scale of (visible sector) spontaneous CP violation to be below that of inflation. However, it must be assessed more carefully in the context of any given model. The explicit CP violation induced by the hidden sector must be large enough to alleviate the domain wall problem, but not so large as to spoil the other goals of the model (e.g., the solution of the Strong CP problem or the suppression of EDMs in a flavor model). If there is a viable parameter space that meets both of these goals, one must assess whether the requisite degree of sequestering can be achieved in a UV completion. The collapsing domain walls formed when the visible sector spontaneously breaks CP could lead to an observable gravitational wave signal. These are clear paths for future work to explore.

\section{Gauge Charge of Parity Domain Walls}
\label{sec:gauge_charge}

The exact stability of parity domain walls and the corresponding conservation of parity domain wall number $\ZZ_2^{\rm PDW}$ appear to be in tension with the expectation that a UV complete theory of quantum gravity cannot admit any global symmetries. Whereas the $(d-2)$-form magnetic symmetry of a discrete internal gauge symmetry (generated by topological Wilson loops) can be broken by dynamical vortices, we have argued in~\S\ref{sec:parityDWQG} that the $(d-2)$-form symmetry $\ZZ_2^{\rm PDW}$ (generated by the topological parity Wilson loop) can never be broken, as parity vortices cannot exist. In this section and the next, we explain how to resolve this tension. The resolution we present is not directly relevant for our phenomenological conclusions, and relies on more mathematics than the previous sections.

Since $\ZZ_2^{\rm PDW}$ is an exact symmetry of any theory of quantum gravity with gauged parity, and can never be broken, there is only one way out. If we truly believe that UV complete quantum gravity does not admit any global symmetries, then we are forced to conclude that $\ZZ_2^{\rm PDW}$ is not a global symmetry, but is actually a gauge symmetry.\footnote{Let us emphasize for clarity that $\ZZ_2^{\rm PDW}$, a $(d-2)$-form symmetry, is not the gauged $0$-form parity symmetry.} This conclusion would provide a natural explanation for the exact stability of parity domain walls. While exactly conserved global charges are forbidden in UV complete quantum gravity, exactly conserved gauge charges are just fine, and are in fact commonplace.

In order to argue that $\ZZ_2^{\rm PDW}$ is gauged, let us recall the difference between global and gauge symmetries from the perspective of quantum gravity. The key distinction is whether or not the conserved charge can be measured remotely. While global charges may be eaten up by black holes, gauge charges radiate gauge flux, which may be measured at a black hole horizon via Gauss's Law. As a result, gauge charges can be conserved in the process of black hole formation and decay \cite{Banks:2010zn}. A related criterion is whether or not we may place a nonzero quantity of charge on a closed manifold: while nonzero global charge may be freely placed on a closed manifold, the net gauge charge on a closed manifold must vanish, again due to Gauss's Law.

\begin{figure}[h]
\centering
\includegraphics [width = 0.65\textwidth]{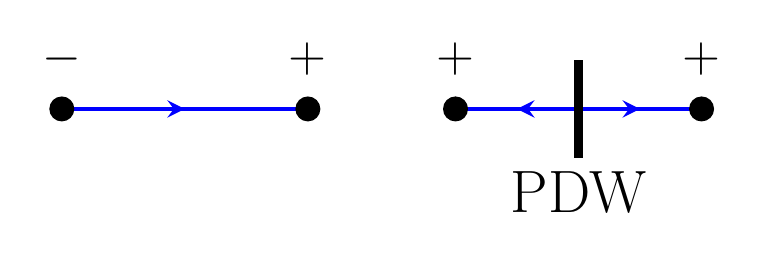}
\caption{
Illustration of the $\ZZ_2^{\rm PDW}$ gauge charge carried by a domain wall, as measured by the orientation outside the domain wall (assuming that the sign of the parity-breaking pseudoscalar is fixed). {\bf Left:} two oppositely oriented points can be connected by an interval with no domain wall. {\bf Right:} a parity domain wall separates two points of the same orientation.
}
\label{fig:oriented_pts}
\end{figure}

Therefore, let us use these criteria to determine whether $\ZZ_2^{\rm PDW}$ is a global or gauge symmetry. First, let us attempt to count the number of parity domain walls (modulo two) inside a region via a measurement at the boundary, to determine if $\ZZ_2^{\rm PDW}$ charge can be remotely measured. Consider a region $R$ of the form
\begin{equation}\label{eq:dw_region}
R = D \times [-L, L],
\end{equation}
containing an unknown number of parity domain walls along the codimension-1 surface $D$.
Let us fix a convention for the sign of the parity breaking condensate $\phi$ outside $R$, say $+ v$ (cf.~\S\ref{subsec:globalparityDW}), allowing us to define an orientation on the boundary $\partial R = D \times \{-L, L\}$. If there is an even number of parity domain walls, then the boundary will consist of two surfaces of opposite orientation, while if there is an odd number, then the boundary will consist of two surfaces of the same orientation (see Fig.~\ref{fig:oriented_pts}). Thus, the topology of the boundary manifold is playing the role of a gauge flux, allowing us to count the number of domain walls inside a region without looking inside. Contrast this with the case of breaking an internal $\ZZ_2$ symmetry: if we fixed the sign of an internal $\ZZ_2$-breaking condensate, there would be nothing left to measure at the boundary. Thus, the first criterion, remote measurability, tells us that $\ZZ_2^{\rm PDW}$ is a gauge symmetry, and its conservation is consistent with black hole (really, black domain wall) physics.

To double check this result, let us turn to our second criterion: whether or not an odd number of parity domain walls may be placed on a closed 1-manifold $C$. However, we now come to a dilemma, as we appear to have previously given two contradictory answers to this question! In~\S\ref{sec:parityDW}, we considered the Klein bottle $K$, and found that the $a$-cycle was a closed 1-manifold transverse to a single parity domain wall, corresponding to a nonzero parity Wilson loop $W_{\rm P}(a) = -1$ (see Fig.~\ref{fig:kleinDW}). On the other hand, our argument for the non-existence of parity vortices was based on the fact that $w_1(TC) = 0$ for every closed 1-manifold $C$, and so the number of parity domain walls ending on any codimension-2 defect encircled by $C$ must be even (see Fig.~\ref{fig:BC} and surrounding discussion).

\begin{figure}[h]
\centering
\includegraphics [width = 0.45\textwidth]{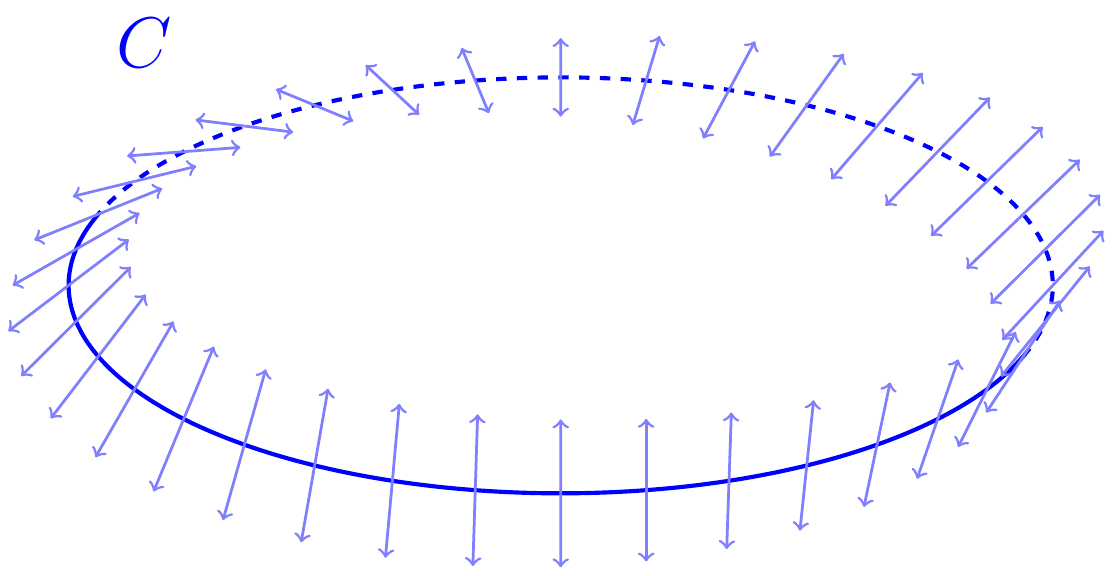}
\caption{
Illustration of a normal bundle $\mathcal{N}_C$ for a 1-manifold $C$ with nonzero parity Wilson loop, $W_{\rm P}(C) = -1$. Note that $C$ is still orientable, even though its normal bundle is not. The total space of this normal bundle is the M\"obius band, a non-orientable 2-manifold.
}
\label{fig:normal}
\end{figure}

Both of these facts are correct: every closed 1-manifold $C$ has $w_1(TC) = 0$, and yet there exist spacetimes with closed 1-manifolds $C$ such that $W_{\rm P}(C) = -1$. The resolution is that the trivial line operator
\begin{equation}\label{gauged_w1_sym_defect}
U(C) = (-1)^{\int_{C} w_1(TC)} = +1,
\end{equation}
which vanishes on every closed 1-manifold, is \emph{not} the parity Wilson loop $W_{\rm P}$. While the operator \eqref{gauged_w1_sym_defect} is defined in terms of the first Stiefel-Whitney class of the 1-manifold $C$, the parity Wilson loop \eqref{def_parity_Wilson} is defined in terms of the first Stiefel-Whitney class of the ambient spacetime $X$. The difference between these two cohomology classes can be quantified by the formula
\begin{equation}\label{eq:normal_w1}
w_1(T X) = w_1(T C) + w_1(\mathcal{N}_{C}),
\end{equation}
along $C$, where $\mathcal{N}_C$ is the \emph{normal bundle} of $C$ in $X$, defined as the complement of $TC$ in $TX$, i.e., $TX = TC \oplus \mathcal{N}_C$ along $C$. Using the formula \eqref{eq:normal_w1} and $w_1(TC) = 0$, we can write the parity Wilson loop entirely in terms of the normal bundle as
\begin{equation}\label{eq:parity_wilson_normal}
W_{\rm P}(C) = (-1)^{\int_{C} w_1(\mathcal{N}_C)}.
\end{equation}
In more physical terms, when one goes around an orientation-reversing curve, it is not the tangent direction to the curve that undergoes a parity transformation, but rather the normal directions, as illustrated in Fig.~\ref{fig:normal}.

Where does this leave us with respect to the question of whether $\ZZ_2^{\rm PDW}$ is a global or gauge symmetry? On the one hand, we have a nontrivial, topological line operator $W_{\rm P}(C)$ that counts parity domain walls, which may be nonzero on a closed 1-manifold $C$, so we might be tempted to say that $\ZZ_2^{\rm PDW}$ is a global charge. This would be a very odd sort of global $(d-2)$-form symmetry: it cannot be broken by \emph{any} codimension-2 defect, and the operator $W_{\rm P}(C)$ is exactly topological (at least up to the quantum gravity cutoff, where the notion of ``line operator'' breaks down). On the other hand, we have argued that this charge is remotely detectable, and the exact conservation of $\ZZ_2^{\rm PDW}$ is consistent with black hole physics, so we would really like to say the symmetry is gauged, despite the existence of a nontrivial topological line operator $W_{\rm P}$.

Our proposal is that a $(d - k - 1)$-form symmetry generated by a nontrivial topological operator $U(M)$ of dimension $k$, which is the identity on all \emph{normally-framed} closed submanifolds, should \emph{not} be considered to be a global symmetry, but should instead be thought of as a gauge symmetry (at least from the perspective that UV complete quantum gravity admits no global symmetries). The condition that $M$ be \emph{normally framed} means that the normal bundle $\mathcal{N}_M$ of $M$ in spacetime is trivial.\footnote{A possible confusion is that ``normal framing'' of a manifold $M$ also sometimes refers to a trivialization of the stable normal bundle, given by $-[TM]$ in KO-theory. Here, we are considering a framing of the relative normal bundle of $M$ in the ambient spacetime $X$, with KO-theory class $[\mathcal{N}_M] = [TX] - [TM]$. In particular, $M$ can have whatever structure our theory allows, and does not need to be a stably framed manifold.} The formula \eqref{eq:parity_wilson_normal}, as well as our prior discussion, show that the parity Wilson loop $W_{\rm P}(C)$ is precisely such an operator, which can only differ from the identity operator on a closed 1-manifold $C$ with a nontrivial normal bundle. In fact, any such $(d - k - 1)$-form symmetry has the following two properties we have already seen for $\ZZ_2^{\rm PDW}$:
\begin{enumerate}

\item The symmetry cannot be broken by any possible defect of codimension $(k+1)$.

\item The associated charge can be measured remotely at the horizon of a codimension-$k$ black brane.

\end{enumerate}
Both of these contexts provide a canonical normal framing, given by the directions tangent to the worldvolume of a codimension-$(k+1)$ defect or a codimension-$k$ black brane, and so in these contexts the symmetry behaves like an ordinary gauge symmetry. As a result, such a symmetry is compatible with black brane physics, and need not (indeed, cannot) be broken in UV complete quantum gravity.

The motivation for our proposal is that while extended operators can be placed on arbitrary submanifolds of spacetime, it really only makes sense to measure charges of quantum states on normally-framed submanifolds. This is most clear for $0$-form symmetries, where the charge should be measured in a Hilbert space $\mathcal{H}_M$ associated to a codimension-1 spatial slice $M$ of spacetime. However, in order to define the Hilbert space $\mathcal{H}_M$, we need to choose an ``arrow of time'' to define a canonical quantization; this arrow of time is precisely a normal framing of $M$ (see~\cite[Section 2.2]{Freed:2016rqq}). If $M$ does not admit a normal framing, then it does not make sense to assign a Hilbert space to $M$. More generally, a fully extended quantum field theory only assigns a $k$-vector space of ``higher quantum states'' to normally-framed submanifolds of codimension $k$~\cite{lurie2008classification}.

For a concrete illustration that it does not make sense to define a Hilbert space on a manifold without a normal framing, consider a 2d string worldsheet theory with gauged parity, i.e., with an orientifold~\cite{Dai:1989ua}. If there were a Hilbert space associated to a circle with nontrivial normal bundle (illustrated in Fig.~\ref{fig:normal}), then the states in this Hilbert space would be twisted sector states for the orientifold, and would correspond to parity vortex operators under the state-operator correspondence. However, as is well-known, there are no twisted sector states for an orientifold~\cite{Polchinski:1996fm},\footnote{It is sometimes said that open strings are the correct notion of twisted sector states of an orientifold. There is a precise sense in which this is true, related to our discussion in~\S\ref{subsec:eotw} regarding the possibility of ``parity vortices'' in the presence of end-of-the-world branes, such as D-branes for the string worldsheet.} which corresponds to the main point of this paper: there is no such thing as a parity vortex operator.

The possibility that a nontrivial topological operator could vanish on all normally-framed submanifolds raises some questions about their role in UV complete quantum gravity. As we have argued, such a topological operator does not generate a global symmetry, and there is no tension between the conservation of the associated charge and black brane physics. Should we think, then, that such a topological operator is allowed in UV complete quantum gravity, contradicting previous claims (including by the authors) that UV complete quantum gravity cannot have any topological operators \cite{Rudelius:2020orz, Heidenreich:2021xpr}?

In the next section, we argue that these sort of topological operators are still forbidden in UV complete quantum gravity, but not because they correspond directly to a global symmetry. Instead, they are a shadow of a new global symmetry that will appear once we break a \emph{different} global symmetry. Thus, in trying to eliminate all global symmetries, we must first break this other global symmetry, causing a new global symmetry to appear, corresponding to the nontrivial topological operator. We then must further break this new global symmetry, to end up with no global symmetries at the end of the day. For the case of the parity Wilson line $W_{\rm P}$, this other symmetry is broken by end-of-the-world branes. In~\S\ref{subsec:eotw}, we show that once end-of-the-world branes are included, there actually \emph{is} a possible object that deserves to be called a parity vortex, in some sense.

\section{Parity Breaking and Cobordism}
\label{sec:cobordism}

In the previous section, we interpreted the exact stability of parity domain walls as the result of an unusual topological gauge charge, whose gauge flux corresponds to the topology of a region's boundary. We further saw that this gauge charge had the unusual property that it need only vanish on a normally-framed closed submanifold, but could be nonzero on a closed submanifold with a nontrivial normal bundle. In this section, we formalize these observations in terms of the well-established mathematical framework of cobordism.

In~\S\ref{subsec:cobordism_intro}, we review the relationship between cobordism groups and topological charges, and explain how the gauge flux radiated by parity domain walls is naturally given by a class in cobordism. In~\S\ref{subsec:adams_ss}, we explain how the Adams Spectral Sequence naturally organizes the possibility that a gauge charge might only vanish on normally framed submanifolds. In~\S\ref{subsec:eotw}, we see how this type of gauge charge corresponds to a new global symmetry that might appear once new defects are added, for the specific case of $\ZZ_2^{\rm PDW}$ in the presence of end-of-the-world branes. In~\S\ref{subsec:orientifolds}, we explain how cobordism provides the correct general notion of ``magnetically charged object'' for spacetime symmetries. We use this understanding to identify further parity defects, beyond domain walls and the non-existent parity vortices, that may exist in a theory with gauged parity symmetry. This section, like the previous one, is not essential to the phenomenological conclusions of this paper.

\subsection{Topological Charges and the Cobordism Conjecture}\label{subsec:cobordism_intro}

\begin{figure}[h]
\centering
\includegraphics [width = 0.8\textwidth]{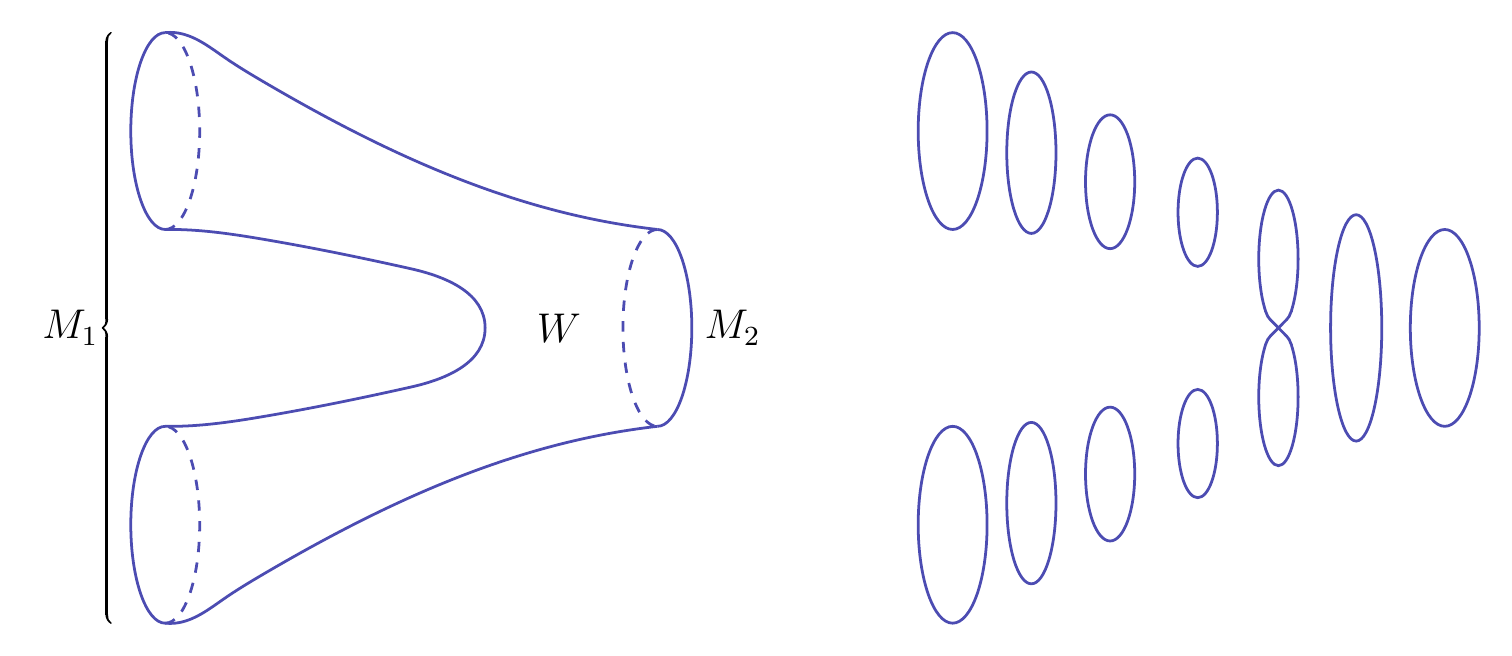}
\caption{
{\bf Left:} Illustration of a cobordism $W$ between the two manifolds $M_1$ and $M_2$.
{\bf Right:} By taking a sequence of slicings through the cobordism, we can view $W$ as the manifold traced out by a topology-changing process.
}
\label{fig:cobordism}
\end{figure}

Cobordism is a mathematically rigorous way of classifying manifolds up to allowed topology changing processes. Given two compact, closed $k$-manifolds $M_1, M_2$, we say that they are \emph{cobordant} if there exists a $(k+1)$-manifold $W$ whose boundary is given by
\begin{equation}
\partial W = \overline M_1 \sqcup M_2,
\end{equation}
where $\overline M_1$ denotes the orientation-reversal of $M_1$. We can view $W$ as the manifold traced out by a process in which we start with $M_1$, perform some topology changes, and arrive at $M_2$, as illustrated in Fig.~\ref{fig:cobordism}. If our manifolds $M_1, M_2$ have some additional structure $\mathcal{X}$, such as an orientation, we may require that this structure is preserved by the process, which means formally that the $\mathcal{X}$-structure extends over $W$ as well. The set of closed $k$-manifolds with $\mathcal{X}$-structure modulo the equivalence relation of $\mathcal{X}$-cobordism forms an abelian group, denoted $\Omega_k^\mathcal{X}$.

In~\cite{McNamara:2019rup}, it was observed that the cobordism group $\Omega_k^\mathcal{X}$ defines a set of charges for a global $(d - k - 1)$-form symmetry in a theory of quantum gravity whose semiclassical description involves a sum over manifolds with $\mathcal{X}$-structure. As UV complete theories of quantum gravity have no global symmetries, it was predicted that the topological symmetries associated to cobordism classes must be either broken or gauged in a UV complete theory. This prediction is known as the Swampland Cobordism Conjecture.\footnote{For a sampling of recent work related to the Cobordism Conjecture, see~\cite{Ooguri:2020sua,Montero:2020icj,Dierigl:2020lai,Buratti:2021yia,Hamada:2021bbz,Buratti:2021fiv,Bedroya:2021fbu,Blumenhagen:2021nmi,Angius:2022aeq,Andriot:2022mri,Blumenhagen:2022mqw,Angius:2022mgh,Blumenhagen:2022bvh,Velazquez:2022eco}.} When these symmetries are broken by dynamical defects, the defects carry a topological gauge charge, and radiate an unusual sort of gauge flux, given by the cobordism class of any manifold with which they link~\cite[\S5.2]{McNamara:2019rup}.

The gauge charge carried by parity domain walls is precisely such a topological gauge charge. Let us consider a gravitational theory with spontaneous parity breaking. Below the scale of parity breaking, such a theory includes a sum over only orientable manifolds, as any non-orientable manifolds must include domain walls (cf.~\S\ref{subsec:globalparityDW}), and are suppressed by the scale of parity breaking. As a result, the relevant cobordism groups below the scale of parity breaking are the oriented cobordism groups $\Omega_k^{SO}$, given in low degree by
\begin{equation}
\Omega_0^{SO} \cong \ZZ, \quad \Omega_1^{SO} = 0, \quad \Omega_2^{SO} = 0, \quad \Omega_3^{SO} = 0.
\end{equation}
In particular, let us consider $\Omega_0^{SO} \cong \ZZ$, the cobordism group of oriented 0-manifolds. A compact, oriented 0-manifold is simply a collection of $n_+$ positively oriented points and $n_-$ negatively oriented points. However, only the difference $n = n_+ - n_-$ is a cobordism invariant, as the oriented interval is a cobordism allowing a positively oriented point and a negatively oriented point to annihilate (illustrated in the left panel of Fig.~\ref{fig:oriented_pts}).

Already, we can see how oriented cobordism captures the gauge flux radiated by a parity domain wall. Given an unknown number of parallel domain walls, we can measure the cobordism class of a 0-manifold linking the domain walls, by fixing the sign of the parity-breaking condensate. In particular, let us choose our 0-manifold to consist of two points, one on each side of the domain walls. There are two possibilities:

\begin{itemize}

\item The two points have opposite orientations. In this case, the cobordism class of the 0-manifold vanishes in $\Omega_0^{SO}$.

\item The two points have the same orientation. In this case, the cobordism class of the 0-manifold is equal to $\pm 2 \in \Omega_0^{SO}$. The sign is not physical, as it may be flipped by changing our convention for the sign of our parity-breaking condensate.

\end{itemize}

\noindent In the second case, the set of two points realizes a nontrivial cobordism class, at least below the scale of parity breaking. Thus, there must be some defect between the two points, along which our description in terms of oriented manifolds must break down. This defect is precisely a parity domain wall, viewed as a defect in the orientation of the spacetime manifold. The parity domain wall breaks the emergent $(d-1)$-form $U(1)$ global symmetry associated to $\Omega_0^{SO} \cong \ZZ$, by connecting two 0-manifolds that were not otherwise cobordant.\footnote{Note that a QFT with a $(d-1)$-form global symmetry is the same thing as a direct sum of QFTs, with kinematically disconnected vacua~\cite{Sharpe:2022ene}. A dynamical domain wall breaks a $(d-1)$-form symmetry by connecting the vacua.} Correspondingly, parity domain walls carry a topological gauge charge, given by the oriented cobordism class of a linked 0-manifold.

For an alternative perspective, we can pass to a description of our theory above the scale of spontaneous parity breaking, where the gravitational path integral includes a sum over both orientable and non-orientable manifolds. Above this scale, the relevant cobordism groups are the unoriented cobordism groups $\Omega_k^O$, given in low degree by
\begin{equation}\label{eq:unoriented_bordism}
\Omega_0^{O} \cong \ZZ_2, \quad \Omega_1^{O} = 0, \quad \Omega_2^{O} \cong \ZZ_2, \quad \Omega_3^{O} = 0.
\end{equation}
While the signed count of points was a cobordism invariant of oriented 0-manifolds, corresponding to $\Omega_0^{SO} \cong \ZZ$, it is no longer a cobordism invariant of unoriented 0-manifolds, as we only have $\Omega_0^O \cong \ZZ_2$. To see this, note that the interval with a single parity domain wall placed in the middle serves as a cobordism trivializing the class of $\pm 2$ oriented points (illustrated in the right panel of Fig.~\ref{fig:oriented_pts}). In terms of symmetries, the $(d-1)$-form $U(1)$ global symmetry associated to $\Omega_0^{SO} \cong \ZZ$ is broken to $\ZZ_2$ once we integrate parity domain walls back into our description. This remaining $(d-1)$-form $\ZZ_2$ global symmetry must be broken in a UV complete theory by an end-of-the-world brane, as we discuss further in~\S\ref{subsec:eotw}.

The next unoriented cobordism group $\Omega_1^O$ also carries relevant information. From \eqref{eq:unoriented_bordism}, we see that $\Omega_1^O$ vanishes. The reason is a basic topological fact discussed previously in \S\ref{sec:parityDWQG}: there is no non-orientable, closed 1-manifold, and thus there is no candidate manifold to realize a nontrivial class in $\Omega_1^O$. We can now rephrase the non-existence of parity vortices in terms of cobordism. Parity vortices are the would-be defects required to break a $(d-2)$-form global symmetry associated to $\Omega_1^O$, but since this group vanishes there is no such global symmetry, and correspondingly no possibility of parity vortices. The vanishing of $\Omega_1^O$ is a further indication that the $(d-2)$-form $\ZZ_2^{\rm PDW}$ symmetry associated to the parity Wilson line $W_{\rm P}$ is gauged. In fact, the closed manifolds $M$ appearing in cobordism groups always carry an implicit normal framing: if $M$ appears as an incoming (outgoing) component of the boundary of a cobordism $W$, the inward (outward) normal vector to the boundary provides a canonical normal framing for $M$ in $W$. Thus, symmetries whose charges vanish on all normally-framed submanifolds do not appear in cobordism groups.

\subsection{Parity Domain Walls and The Adams Spectral Sequence}\label{subsec:adams_ss}

We have seen that the gauge flux radiated by parity domain walls is given by the class in oriented cobordism $\Omega_0^{SO}$ of a 0-manifold linking their worldvolume. This description is appropriate for physics below the scale of parity breaking. Above the scale of parity breaking, we saw that the gauging of $\ZZ_2^{\rm PDW}$ corresponds to the vanishing of the unoriented cobordism group $\Omega_1^O$, despite the existence of the nontrivial topological line operator $W_{\rm P}$, associated to the nontrivial characteristic class $w_1$. In this section, we connect these descriptions above and below the scale of parity breaking via the Adams Spectral Sequence. To motivate the Adams Spectral Sequence, let us generalize the phenomenon we seek to understand: that a nontrivial characteristic class, like $w_1$, might vanish on all normally-framed submanifolds, and thus fail to produce a nontrivial cobordism invariant.

Let us fix some structure $\mathcal{X}$, and consider the set of all characteristic classes of $\mathcal{X}$-manifolds. In general, a characteristic class $a$ of degree $k$ defines a potential cobordism class in $\Omega_k^\mathcal{X}$. The only way the characteristic class $a$ could fail to define a nontrivial cobordism invariant in $\Omega_k^{\mathcal{X}}$ is if the characteristic class vanishes on all normally framed, closed $\mathcal{X}$-manifolds of dimension $k$. When this happens, the characteristic class $a$ corresponds to some closed $k$-manifold $M$ with a nontrivial normal bundle $\mathcal{N}_M$. For the case of $w_1$, this manifold is illustrated in Fig.~\ref{fig:normal}. Alternatively, we can cut out an open submanifold $U$ of $M$ so that the normal bundle is trivial over $M \setminus U$, and say that $a$ corresponds to $M \setminus U$, a normally-framed $k$-manifold with boundary $\partial U$. The $(k-1)$-dimensional boundary $\partial U$ corresponds to another potential cobordism class $b$ in $\Omega_{k - 1}^\mathcal{X}$, which is trivialized by the cobordism $M \setminus U: \partial U \to \varnothing$. For $w_1$, the manifold with boundary $M \setminus U$ is illustrated in the right hand panel of Fig.~\ref{fig:eotw}, and the boundary $\partial U$ consists of two points of the same orientation. 

The Adams Spectral Sequence formalizes this possibility, and provides a way to compute cobordism groups given characteristic classes.\footnote{For a practical introduction to the Adams Spectral Sequence, see \cite{beaudry2018guide}. For more formal textbook accounts, see \cite{kochman1996bordism, ravenel2003complex}.} The Adams Spectral Sequence consists of a sequence of ``pages'' $E_r$, each of which contains many potential cobordism classes. These potential cobordism classes are acted on by ``differentials'' $d_r$, which encode the possibility that one potential class $a$, of dimension $k$, can correspond to a normally framed manifold with boundary. The boundary then corresponds to another potential class $b$, of dimension $(k-1)$, as described above. This simultaneously trivializes the class of $b$, breaking the corresponding symmetry, while also assigning a topological gauge charge to the object $a$. In the subsequent page $E_{r+1}$, both the classes $a$ and $b$ connected by the differential $d_r$ are removed, and a new differential $d_{r+1}$ acts on the remaining potential classes.\footnote{The need for a new differential may be understood as follows. Suppose we have two $k$-manifolds $M_i, i = 1, 2$, with nontrivial normal bundles $\mathcal{N}_{M_i}$, so that we may cut out open sets $U_i$ from $M_i$ such that the normal bundles are trivial over $M_i \setminus U_i$, and such that the boundaries are related by $\partial U_1 = \overline{\partial U_2}$. We may glue the two boundaries together, to form a closed $k$-manifold $\widetilde M$. However, while the normal bundles are trivialized over $M_i \setminus U_i$, the trivializations may not agree on the boundaries $\partial U_i$, and so the normal bundle $\mathcal{N}_{\widetilde M}$ may still be nontrivial, though only along a locus of higher codimension.} The limiting page $E_\infty$ corresponds to the actual cobordism groups, up to possible group extensions.

\begin{figure}[h]
\centering
\vspace{-12pt}
\includegraphics [width = 0.5\textwidth]{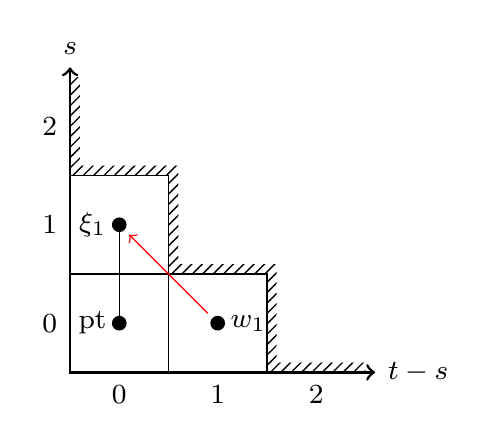}
\vspace{-12pt}
\caption{
A small piece of the $E_1$ page of the Adams Spectral Sequence for unoriented cobordism $\Omega_{t - s}^O$ at the prime $2$. The horizontal axis corresponds to the dimension of the manifold, while the vertical axis is the Adams filtration. The vertical line connecting the class of the point to $\xi_1$ indicates multiplication by $2$. There is a nonzero differential $d_1$, drawn in red, connecting $w_1$ to $\xi_1$. This differential encodes the fact that the ``parity domain wall number'' symmetry with current $w_1$ is gauged, with the gauge flux given by $\xi_1$, the class of two points of the same orientation.
}
\label{fig:adams_ss}
\end{figure}

In terms of the Adams Spectral Sequence, we can succinctly describe the gauging of $\ZZ_2^{\rm PDW}$ as the following mathematical fact: there is a differential on the $E_1$ page of the Adams Spectral Sequence for unoriented cobordism, connecting the potential class $w_1$ to the potential class of two points of the same orientation. This differential is illustrated in Fig.~\ref{fig:adams_ss}. The $E_1$ page of the Adams Spectral Sequence may be less familiar than the $E_2$ page. While the $E_2$ page and onwards depend only on the set of characteristic classes as a module over the Steenrod algebra $\mathcal{A}$, the $E_1$ page depends on the choice of Adams resolution. A universal choice is the cobar complex, formed by taking iterated tensor products with the dual Steenrod algebra $\mathcal{A}^\vee$. At the prime $2$, the dual Steenrod algebra is given by $\mathcal{A}^\vee = \ZZ_2[\xi_1, \xi_2, \dots]$, a polynomial algebra on elements $\xi_i$ in degree $2^i - 1$. The element $\xi_1$ is the potential cobordism class of two points of the same orientation.

The differential illustrated in Fig.~\ref{fig:adams_ss} is merely the first of an infinite number of nonzero differentials on the $E_1$ page of the Adams spectral sequence for unoriented cobordism. Correspondingly, the vanishing of $w_1$ on all normally-framed closed 1-manifolds is the first in a set of infinitely many \emph{Wu relations}, polynomials of degree $k$ in the Stiefel-Whitney classes that vanish on all normally-framed closed $k$-manifolds. Our discussion suggests that the Wu relations, as well as similar relations for other choices of structure, should all be viewed as gauge constraints in quantum gravity. All of these gauge constraints are of the form discussed in~\S\ref{sec:gauge_charge}, and are associated with nontrivial topological symmetry operators which vanish along all normally framed submanifolds.

We now have the machinery to address the question raised at the end of~\S\ref{sec:gauge_charge}, as to whether this sort of nontrivial topological operator is allowed in UV complete quantum gravity. These operators correspond to characteristic classes that are eliminated via a differential in the Adams Spectral Sequence, as opposed to characteristic classes that vanish identically, which are gauged in the usual sense and do not correspond to any topological operator.\footnote{Contrast the case of $w_1$ in unoriented and oriented bordism: while $W_{\rm P}(C)$ vanishes for any normally-framed curve in a non-orientable manifold, it vanishes for any curve whatsoever in an orientable manifold, by definition.} Naively, the Cobordism Conjecture seems compatible with nontrivial characteristic classes, so long as every characteristic class is eliminated by a differential in the Adams spectral sequence. However, we can actually prove that if this happens, then there were no nontrivial characteristic classes to begin with. In other words, if the Cobordism Conjecture holds and \emph{all} cobordism groups vanish, then there cannot be any topological operators whatsoever, even ones which vanish on normally-framed manifolds.

To prove that trivial cobordism groups for some structure $\mathcal{X}$ imply the absence of any nontrivial $\mathcal{X}$-characteristic classes, we need to use a bit more algebraic topology. Consider the Madsen-Tillmann cobordism spectrum $M T \mathcal{X}$ (see~\cite{Freed:2016rqq} for a definition), which has the property that its homotopy groups are $\mathcal{X}$-cobordism groups:
\begin{equation}\label{eq:pik_cobordism}
\pi_k (MT\mathcal{X}) = \Omega_k^\mathcal{X}.
\end{equation}
Let us recall that characteristic classes of $\mathcal{X}$-manifolds (valued in $\ZZ_2$) are given by the $\ZZ_2$-cohomology $H^*(MT\mathcal{X}; \ZZ_2)$.\footnote{More generally, $\mathcal{X}$-characteristic classes valued in a generalized cohomology theory $E$ are given by the $E$-cohomology $E^*(MT \mathcal{X})$. Our argument carries over verbatim for $E$-valued characteristic classes. This sort of generalized characteristic class was recently studied in the context of the Cobordism Conjecture in~\cite{Blumenhagen:2022bvh,Blumenhagen:2021nmi}.} Now, suppose $\mathcal{X}$ satisfies the Cobordism Conjecture, in that $\Omega_k^\mathcal{X} = 0$ for all $k$.\footnote{This will likely require the inclusion of singularities, and potentially more exotic spaces, in our notion of structure $\mathcal{X}$, which is a bit broader than what mathematicians normally consider. In any case, the logic is the same.} By \eqref{eq:pik_cobordism}, this is equivalent to saying that $MT\mathcal{X}$ is contractible. But a contractible spectrum has trivial cohomology, and so there can be no nontrivial characteristic classes of $\mathcal{X}$-manifolds.

This statement, that $\pi_k(MT\mathcal{X})$ vanishing for all $k$ implies that $H^k(MT\mathcal{X}; \ZZ_2)$ vanishes for all $k$, does not hold dimension-by-dimension. Indeed, we have seen that $\pi_1(MTO) = \Omega_1^O = 0$, while $H^1(MTO; \ZZ_2) = \ZZ_2$, generated by $w_1$. Instead, this nonzero cohomology group is related to the nontrivial cobordism group $\pi_0(MTO) = \Omega_0^O = \ZZ_2$, and the corresponding characteristic class $w_0 \in H^0(MTO; \ZZ_2)$, via the action of a Steenrod square,
\begin{equation}
{\rm Sq}^1: H^0(MTO; \ZZ_2) \to H^1(MTO; \ZZ_2), \quad w_0 \mapsto w_1.
\end{equation}
This Steenrod square is precisely responsible for the nonzero differential depicted in Fig.~\ref{fig:adams_ss}. The cobordism group $\Omega_0^O = \ZZ_2$ corrresponds to the remaining $(d-1)$-form symmetry remaining after the $U(1)$ symmetry corresponding to $\Omega_0^{SO} = \ZZ$ is broken by parity domain walls. As mentioned in \S\ref{subsec:cobordism_intro}, this symmetry must be broken in a UV complete theory by the existence of end-of-the-world branes.

\subsection{End-of-the-World Branes and Parity Vortices}\label{subsec:eotw}

Let us see what happens once end-of-the-world branes are included. As these are end-of-the-world branes for the parity-breaking phase, their worldvolume theory need not be parity-symmetric, and so they should be defined on oriented manifolds. To describe the cobordism of a theory with dynamical defects, we must consider cobordism of manifolds with singularities (also known as the Baas-Sullivan construction~\cite{baas,sullivan1971geometric}). For end-of-the-world branes, these singularities are given by a nontrivial boundary, equipped with an orientation, on which the end-of-the-world brane lives.

\begin{figure}[h]
\centering
\includegraphics [width = 0.4\textwidth]{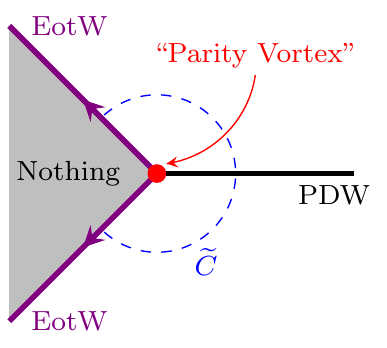}\quad\quad\quad\includegraphics [width = 0.45\textwidth]{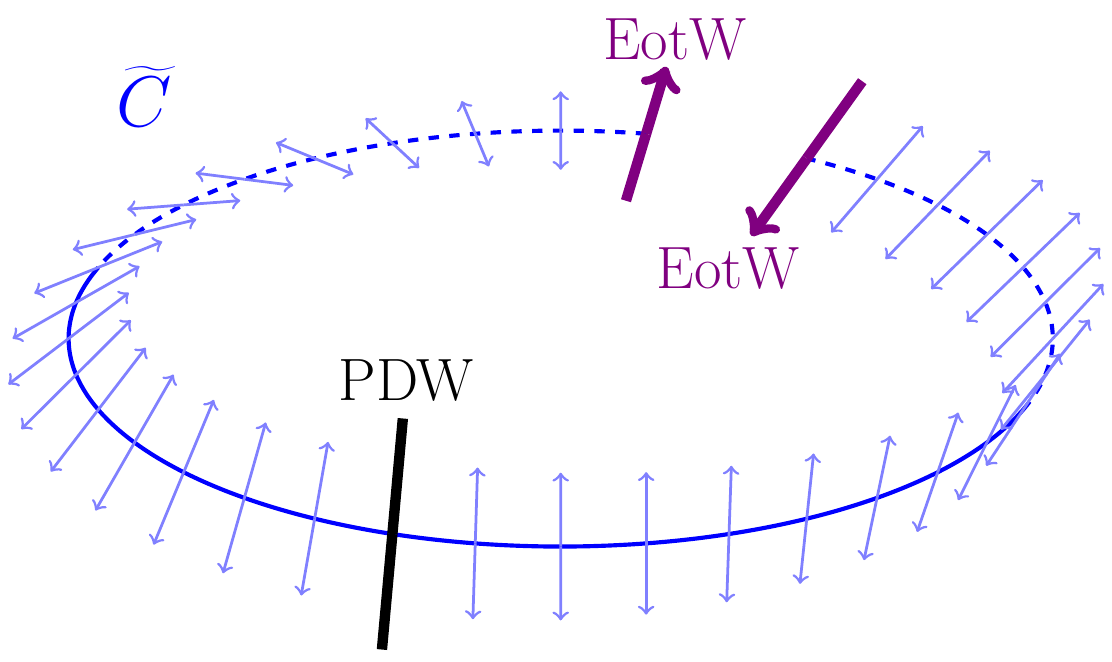}
\caption{
With dynamical end-of-the-world branes, a notion of ``parity vortex'' appears.
{\bf Left:} A parity domain wall (black line) can end at a junction (red dot) with two end-of-the-world branes (purple lines). The two end-of-the-world branes are both oriented outwards from the junction. The left triangular region (grey) is a bubble of nothing~\cite{Witten:1981gj}. The ``parity vortex'' is encircled by a 1-manifold with boundary $\widetilde C$ (dashed blue curve), which represents a nontrivial class in cobordism of manifolds with end-of-the-world branes.
{\bf Right:} Another view of $\widetilde C$, emphasizing the normal bundle. If we wanted to annihilate the two end-of-the-world branes, to form a closed 1-manifold, we would need to introduce a half-twist in the normal direction, obtaining the manifold $C$ depicted in Fig.~\ref{fig:normal}. Despite this twist, $\widetilde C$ is still normally framed, but the normal framing does not extend to $C$. The manifold $\widetilde C$ is diffeomorphic to the right hand panel of Fig.~\ref{fig:oriented_pts}, with end-of-the-world branes placed at the boundaries.
}
\label{fig:eotw}
\end{figure}

In this cobordism theory, a new 1-dimensional cobordism class appears, given by an interval whose boundary consists of two points of the same orientation.\footnote{This class also appears in string theory, as the class of M-theory on $S^1/\ZZ_2$. The two boundaries, each carrying a 10d $\mathcal{N} = (1, 0)$ $E_8$ gauge theory, have the same chirality. This construction gives the Ho\v{r}ava-Witten description of heterotic $E_8 \times E_8$ string theory~\cite{Horava:1995qa,Horava:1996ma}.} There is a single parity domain wall in the middle of the interval, and an end-of-the-world brane on each of the two endpoints. We will denote this manifold by $\widetilde C$; it is illustrated in Fig.~\ref{fig:eotw}. Let us note that $\widetilde C$ is diffeomorphic to the manifold with boundary used to measure the gauge flux radiated by parity domain walls, illustrated in the right hand panel of Fig.~\ref{fig:oriented_pts}.

This new cobordism class $[\widetilde C]$ corresponds to a new $(d-2)$-form global symmetry. This new global symmetry is directly related to the parity Wilson loop $W_{\rm P}$. While $W_{\rm P}$ vanishes on any normally framed, closed 1-manifold, it does \emph{not} vanish on the normally framed 1-manifold with boundary $\widetilde C$. This can be seen from the fact that $\widetilde C$ contains just a single parity domain wall, and so the net $\ZZ_2^{\rm PDW}$ charge on $\widetilde C$ does not vanish. Somehow, by adding end-of-the-world domain walls, we seem to have \emph{ungauged} $\ZZ_2^{\rm PDW}$! This new $(d-2)$-form global symmetry must be broken by some additional defect of codimension-2 in spacetime. As this defect serves as the endpoint of a single parity domain wall, it deserves, in some sense, to be called a ``parity vortex.'' This perspective is reinforced if we try to annihilate the two end-of-the-world branes at the boundary of $\widetilde C$, as depicted in the right panel of Fig.~\ref{fig:eotw}. Since the two end-of-the-world branes have the same orientation, we must introduce a normal half-twist in order to annihilate them, and so we would obtain the closed manifold $C$ illustrated in Fig.~\ref{fig:normal}, with a nontrivial normal bundle.

The situation becomes much less mysterious if we think a bit more carefully about domain wall charges in this setup. In addition to the $(d-2)$-form $\ZZ_2^{\rm PDW}$ symmetry, there is also a $(d-2)$-form $U(1)^{\rm EotW}$ symmetry, corresponding to the conservation of end-of-the-world brane number.\footnote{This is a $U(1)$ symmetry, because the end-of-the-world branes are oriented, and have an integer conserved charge.} However, there are discrete gauge constraints on a normally-framed 1-manifold. If we let $n_{\rm EotW}, n_{\rm PDW}$ denote the number of end-of-the-world branes and parity domain walls, respectively, we must have
\begin{equation}\label{eq:discrete_gauge_constraint}
n_{\rm EotW} = 0 \mod 2, \quad n_{\rm PDW} = \frac{n_{\rm EotW}}{2} \mod 2.
\end{equation}
This holds on $\widetilde C$, since we have $n_{\rm EotW} = 2, n_{\rm PDW} = 1$. When there are no end-of-the-world branes ($n_{\rm EotW} = 0$), this gauge constraint reduces to our previous gauge constraint, that there must be an even number of parity domain walls on a normally-framed, closed 1-manifold. With the discrete gauge constraint~\eqref{eq:discrete_gauge_constraint}, the actual global $(d-2)$-form symmetry is
\begin{equation}\label{eq:remaining_global_sym}
\frac{\ZZ_2^{\rm PDW} \times U(1)^{\rm EotW}}{\ZZ_4} \cong U(1).
\end{equation}
What has happened is not that $\ZZ_2^{\rm PDW}$ has been ``ungauged,'' but rather that there are now two $(d-2)$-form symmetries, and a diagonal combination has been gauged. In other words, the gauge charge of a parity domain wall is also carried by a pair of end-of-the-world branes. This can be seen in terms of the gauge flux, as both radiate the cobordism class of two oriented points (cf.~\S\ref{subsec:cobordism_intro}).

In these terms, the new codimension-2 defect required to break the remaining $(d-2)$-form global symmetry~\eqref{eq:remaining_global_sym} associated with the cobordism class $[\widetilde C]$, which we had been calling a ``parity vortex,'' gets a different interpretation. It is a junction of a parity domain wall and a pair of end-of-the-world branes, as illustrated in the left hand panel of Fig.~\ref{fig:eotw}.\footnote{This junction can also be viewed as a parity domain wall for the worldvolume theory on the end-of-the-world brane. The bulk parity domain wall ending on the junction is then a parity domain wall for the coupled bulk-boundary system.} This junction allows a parity domain wall to dynamically turn into a pair of end-of-the-world branes, conserving gauge charge while breaking the global symmetry~\eqref{eq:remaining_global_sym}. The region between the two end-of-the-world branes is a bubble of nothing~\cite{Witten:1981gj}: a void where spacetime does not exist. Whether one should call this junction a ``parity vortex'' is a matter of taste. It is a very different object from the ``parity vortices'' we ruled out in~\S\ref{sec:parityDWQG}.

In particular, this junction does not spoil our main conclusion, that parity domain walls formed in a cosmological context would be exactly stable. In the theories we are interested in, parity is spontaneously broken at low energies by conventional quantum field theory dynamics. In contrast, an end-of-the-world brane is an intrinsically quantum-gravitational object, whose tension should generically be related to the quantum gravity cutoff energy.  Thus, we expect it to be energetically impossible for a parity domain wall to decay to a pair of end-of-the-world branes. Moreover, if such a process were possible, it would create a bubble-of-nothing instability~\cite{Witten:1981gj, GarciaEtxebarria:2020xsr} that would destroy the ambient spacetime, far from any desired process that removes domain walls and restores a conventional cosmology.

\subsection{Orientifold Planes and I-folds}\label{subsec:orientifolds}

In general, a UV complete theory of quantum gravity is expected to have a complete spectrum of electrically and magnetically charged objects~\cite{Polchinski:2003bq}. As discussed in~\S\ref{subsec:twistvortex}, the magnetically charged objects of an internal $\ZZ_2$ gauge symmetry are codimension-2 vortices. However, we have seen that gauged parity is qualitatively different: there is no such thing as a parity vortex (at least without end-of-the-world branes), and one may wonder whether magnetic completeness has any meaning for gauged parity symmetry. The answer is that it does have meaning, but the proper notion of ``magnetically charged object'' for spacetime symmetries is given by cobordism. While the cobordism group $\Omega_1^O$ vanishes, and there is no possibility of parity vortices, the higher cobordism groups $\Omega_k^O$ do not vanish in general, and require a rich spectrum of magnetically charged objects under parity. Many of these objects have familiar incarnations in string theory as orientifold planes.

To illustrate this notion of magnetically charged object, let us consider the cobordism group $\Omega_2^O = \ZZ_2$, as given in \eqref{eq:unoriented_bordism}. The nonzero element in $\Omega_2^O$ is given by the cobordism class of the real projective plane $\RR \mathbb{P}^2 = S^2 / \ZZ_2$, and is associated to the second Stiefel-Whitney class $w_2$. In a UV complete theory of quantum gravity with gauged parity symmetry, the $(d-3)$-form symmetry with current $w_2$ must either be gauged or broken. If it is gauged, then the theory cannot be placed on the manifold $\RR \mathbb{P}^2$. One way this can happen is if the theory contains fermions, and the parity symmetry acts on the fermions in a way that requires pin$^+$ structure,\footnote{The (broken) CP symmetry of the Standard Model, acting on Weyl fermions by $\psi_\alpha(t, \vec x) \mapsto i \psi^{\dagger \dot \alpha}(t, - \vec x)$, requires pin$^+$ structure (see \cite{Wang:2021nmi} for a review). However, if there is an additional $\ZZ_4$ symmetry under which all fermions have odd charge and all bosons have even charge, we may twist the CP symmetry by this $\ZZ_4$ symmetry to obtain a pin$^-$ symmetry, which acts on the Weyl fermions by $\psi_\alpha(t, \vec x) \mapsto \pm \psi^{\dagger \dot \alpha}(t, - \vec x)$, with signs chosen to preserve all fermion mass terms. This can happen if there are right-handed neutrinos with only Dirac masses, in which case we may take the $\ZZ_4$ subgroup of the non-anomalous $U(1)_{B - L}$ symmetry~\cite{Wilczek:1979et, Krauss:1988zc, Garcia-Etxebarria:2018ajm, Wang:2020xyo}. On the other hand, Majorana masses are incompatible with a pin$^-$ symmetry~\cite{Berg:2000ne}, though they could be generated below the scale of spontaneous $\ZZ_4$ breaking. It would be interesting to explore the phenomenological consequences of the various possible combinations of CP symmetry with fermion number in the Standard Model.} as $\RR \mathbb{P}^2$ is not a pin$^+$ manifold.

\begin{figure}[h]
\centering
\includegraphics [width = 0.6\textwidth]{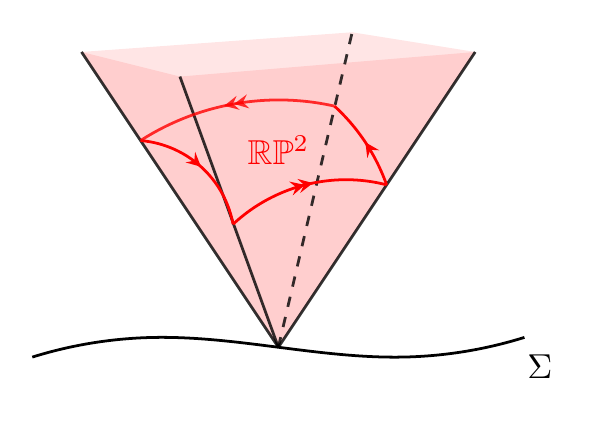}
\caption{
Illustration of a conical defect of codimension 3 in spacetime, with local topology $\RR^3/\ZZ_2$. Such a defect, sometimes called an ``I-fold,'' lives along a codimension-3 locus $\Sigma$, and is linked by the manifold $\RR \mathbb{P}^2$, corresponding to the nontrivial cobordism class in $\Omega_2^O = \ZZ_2$. Such defects must exist in UV complete theories of quantum gravity that can be defined on the manifold $\RR \mathbb{P}^2$. For example, in Type IIA string theory, this defect is the O6-plane.
}
\label{fig:rp2_defect}
\end{figure}

On the other hand, if the theory does not have fermions, or the parity symmetry acts on the fermions in a way that requires pin$^-$ structure, then the $(d-3)$-form symmetry with current $w_2$ must be broken by defects of codimension 3 in spacetime (particles in 4d). This is in contrast to the case of $w_1$, which we have argued is always automatically gauged. Viewed from far away, a defect that breaks this $(d-3)$-form symmetry looks like a cone over $\RR \mathbb{P}^2$, with topology $\RR^3/\ZZ_2$, as illustrated in Fig.~\ref{fig:rp2_defect}. This defect corresponds to a codimension-3 boundary condition on the gravitational path integral, defined by cutting out a codimension-3 locus $\Sigma$ from spacetime, and summing over spacetimes with boundary $\Sigma \times \RR \mathbb{P}^2$ (cf.~\S\ref{sec:parityDWQG}). In Type IIA string theory, this defect is given by a well known object: the O6-plane. More generally, such objects have been called ``I-folds,'' for ``inversion-fold,'' and have played a key role in recent work reconstructing the known string landscape from swampland principles in $d > 6$ supergravity~\cite{Montero:2020icj}. In fact, we have seen another example of an I-fold already, with topology $\RR/\ZZ_2$: the end-of-the-world branes considered in~\S\ref{subsec:eotw}. I-folds are the proper notion of magnetically charged object in a theory with gauged parity, rather than the non-existent parity strings.

\begin{figure}[h]
\centering
\includegraphics [width = 0.45\textwidth]{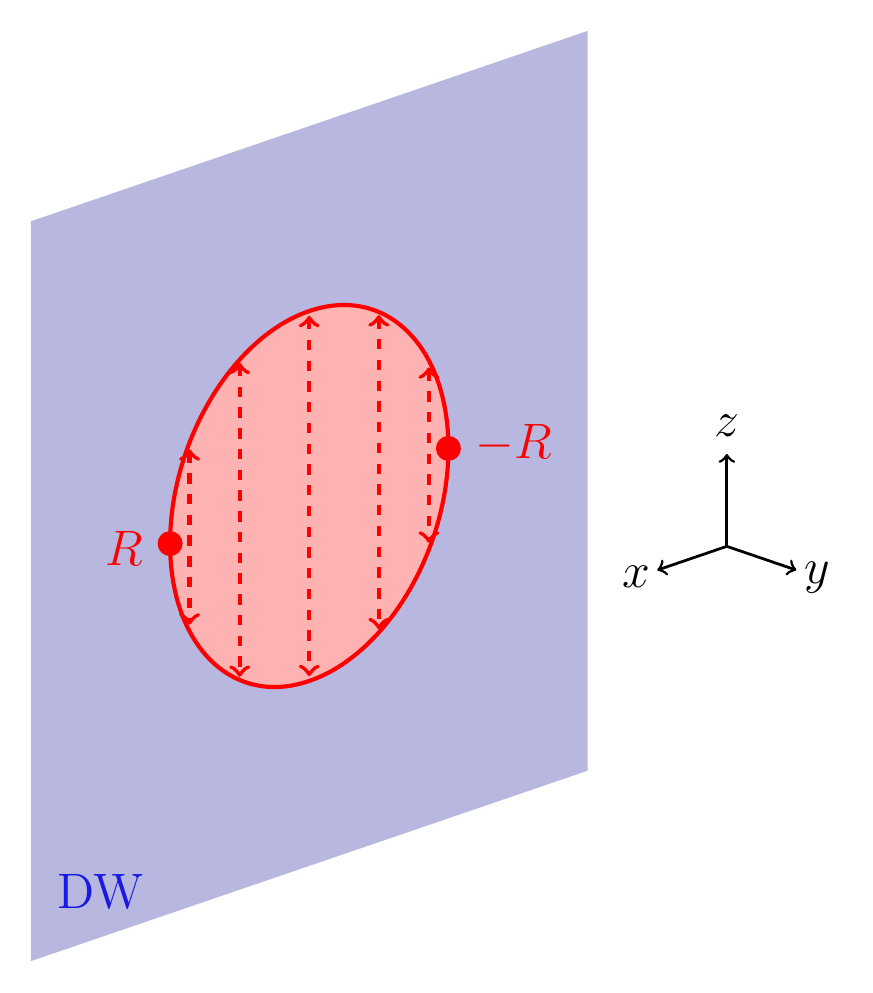}
\caption{
Cutting open of a disk (red shaded region) within a domain wall (blue region). The front and back faces of the disk are identified after a reflection in the $z$-axis (red dashed arrows), as in~\eqref{eq:cutting_and_gluing}, so that an observer passing through the disk will be flipped vertically. The red dots at $x = \pm R$ indicate I-folds. The red circle is not a boundary of the domain wall, because each point there is identified with a point on the opposite ``edge.'' Because the cutting and gluing operation destroys entanglement of local quantum fields, the disk is a firewall.
}
\label{fig:Ifold_hole}
\end{figure}

A reasonable worry is that dynamical I-folds could spoil our main arguments by providing a decay channel for parity domain walls. The upshot is that they do not, but to see this is rather subtle, as I-folds seem to allow a hole to open in the middle of a parity domain wall. For concreteness, let us work in 4d, where the I-fold $\RR^3/\ZZ_2$ is a particle. The nucleation of a pair of I-folds in the middle of a parity domain wall can produce a region with a modified transition function, as depicted in Fig.~\ref{fig:Ifold_hole}. Spacetime is cut open along the disk of radius $R$ in the $(x-z)$-plane at $y = 0$, and glued back together via the identification
\begin{equation}\label{eq:cutting_and_gluing}
(x, +\epsilon, z) \sim (x, -\epsilon, -z), \quad x^2 + z^2 \leq R^2.
\end{equation}
We take the short-distance cutoff $\epsilon$ to be the inverse of the quantum gravity cutoff, $\epsilon \sim \Lambda_{\rm QG}^{-1}$. There are two I-folds, located at the two points $(\pm R, 0, 0)$. This process appears to provide the type of decay channel we attempted to rule out in~\S\ref{sec:parityDWQG}. To argue that this process does not destabilize the domain wall, we have to argue energetically. Recall that the nucleation of a hole in a symmetry-breaking domain wall is a tunneling process, and involves a competition between an energy cost to nucleate a loop of string and an energy benefit to destroy the domain wall. The energy cost of the loop of string scales with the perimeter of the hole, while the energy benefit scales with the area, and so above a critical radius it becomes energetically favorable to nucleate such a hole, which will expand and destroy the domain wall.

This is not so for the hole produced in a parity domain wall by nucleation of I-folds. In addition to the energy benefit from destroying the domain wall, associated with the zero-mode of a parity-breaking condensate, there is an additional energy cost from cutting and gluing spacetime at a macroscopic scale as in \eqref{eq:cutting_and_gluing}, associated with the short-distance correlations of fluctuations of quantum fields~\cite{Bisognano:1975ih, Hawking:1975vcx}. In other words, the supposed ``hole'' in the parity domain wall is no hole, but is instead a firewall! The tension of this firewall scales as $T_{\rm FW} \sim N_{\rm dof} \epsilon^{-3} \sim N_{\rm dof} \Lambda_{\rm QG}^3 \sim M_\mathrm{Pl}^2 \Lambda_{\rm QG}$ in 4d~\cite[\S3.5]{Harlow:2014yka}, where $N_{\rm dof} \sim M_\mathrm{Pl}^2/\Lambda_{\rm QG}^2$ is the number of light dynamical degrees of freedom~\cite{Veneziano:2001ah}. This produces an enormous area-scaling energy cost that will always beat out the energy benefit of destroying the domain wall. This sort of wildly non-local process stands in contrast to what would have been possible with parity vortices, which would have provided a local boundary condition for parity domain walls, had they existed. Note that the domain wall in Fig.~\ref{fig:Ifold_hole} does not actually end anywhere; the two ``edges'' of the domain wall along the top and bottom semicircles of radius $R$ are glued to each other by \eqref{eq:cutting_and_gluing}.

\begin{figure}[h]
\centering
\includegraphics [width = 0.65\textwidth]{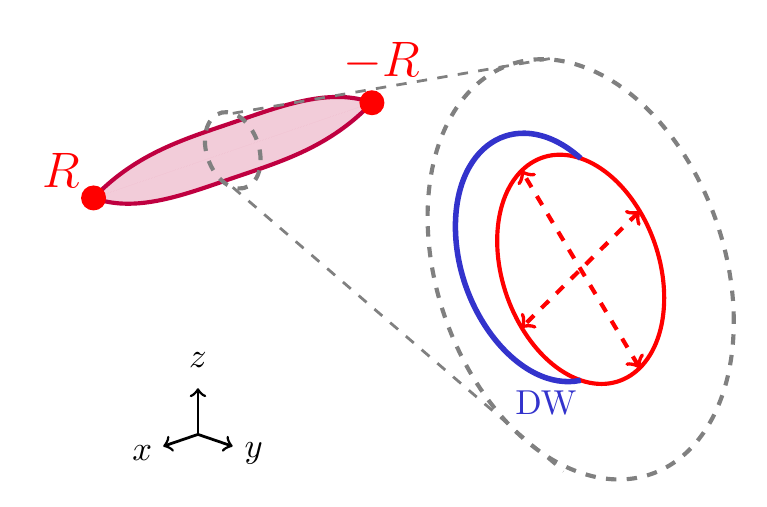}
\caption{
In a phase with broken parity, two I-folds (located at $x = \pm R$) are confined by an $\RR \mathbb{P}^2$-string (shaded region). A cross section of this string (shown within the dashed grey circle) consists of a circle (red) with a crosscap identification of opposite points (red dashed arrows), as in~\eqref{eq:cross_cap_string}, together with a domain wall surface (blue). The two ``ends'' of the domain wall in the figure are identified points, so the domain wall has no boundary.
}
\label{fig:Ifold_string}
\end{figure}

By a similar argument, we can see that I-folds are confined in the parity-breaking phase. In some sense, this is to be expected: I-folds are magnetically charged objects under gauged parity, and a parity-breaking condensate higgses the gauged parity symmetry, so it is natural to expect I-folds to confine. Let us argue in 4d. Consider the process of nucleating a pair of I-folds away from any parity domain walls. To avoid firewalls, we must avoid any macroscopic cutting-and-gluing of spacetime, and so we expect the modification of the spacetime manifold to be localized to a string connecting the two I-folds, as in Fig.~\ref{fig:Ifold_string}. Spacetime is cut open along an interval in the $x$-axis between the I-folds at $(\pm R, 0, 0)$, and glued together via
\begin{equation}\label{eq:cross_cap_string}
(x, + \epsilon \cos \theta, + \epsilon \sin \theta) \sim (x, - \epsilon \cos \theta, - \epsilon \sin \theta), \quad x \in [-R, R],
\end{equation}
where $\theta$ denotes the azimuthal angle around the string. Infinitesimally close points remain infinitesimally close after the identification~\eqref{eq:cross_cap_string}, so there is no firewall. This string is a gravitational soliton with cross-sectional topology $\RR \mathbb{P}^2$ \cite{McNamara:2019rup, McNamara:2021cuo}, and so we call it the ``$\RR \mathbb{P}^2$-string.'' As $\RR \mathbb{P}^2$ is non-orientable, there must be a tube of parity domain wall along this string, corresponding to the non-vanishing of $w_1(\RR \mathbb{P}^2)$. We may give a rough estimate the tension of the $\RR \mathbb{P}^2$-string in terms of the domain wall tension $T_{\rm DW}$ as $T_{\rm string} \sim T_{\rm DW} \epsilon \sim T_{\rm DW} \Lambda_{\rm QG}^{-1}$, as the inverse quantum gravity cutoff $\epsilon \sim \Lambda_{\rm QG}^{-1}$ is the natural size of gravitational solitons~\cite{Hawking:1978pog}.\footnote{Reference~\cite{Hawking:1978pog} actually gives the Planck length as the natural size of gravitational solitons. However, this is the Planck length in a UV description of the gravitational path integral, while $M_\mathrm{Pl}^{-1}$ is the Planck length as read off from the IR effective action. The inverse quantum gravity cutoff $\Lambda_{\rm QG}^{-1}$ may be viewed as the UV value of the Planck length. For example, in models with extra dimensions, $\Lambda_{\rm QG}^{-1}$ is the higher-dimensional Planck length.} This tensionful $\RR \mathbb{P}^2$-string confines the I-folds in the parity-breaking phase, but they may become deconfined above the scale of parity breaking.\footnote{For instance, O6-planes are not confined in Type IIA string theory.} This may be seen via cobordism: the class $[\RR \mathbb{P}^2]$ does not exist in oriented bordism, and so we should not expect to find isolated I-folds below the scale of parity breaking.

Be careful not to confuse the solitonic $\RR \mathbb{P}^2$-string with the non-existent parity string (as we did, originally). While a parity string would be detectable from far away due to an ``Aharonov-Bohm'' parity reflection around it, the spacetime around the $\RR \mathbb{P}^2$-string is perfectly orientable, and so the $\RR \mathbb{P}^2$-string cannot be remotely detected. 

\section{Conclusions}
\label{sec:conclusions}

Many current experiments search for indirect evidence of new spacetime symmetries that are spontaneously broken in the world around us. Supersymmetry and extra dimensions are two examples, both major paradigms for possible extensions of the Standard Model. Discrete, orientation-reversing spacetime symmetries, like parity or CP, receive less attention because experiment has already revealed them to be maximally broken. Nonetheless, it remains a possibility that they are fundamental gauge symmetries in quantum gravity, broken only spontaneously, perhaps at energies far below the Planck scale. We should understand possible experimental signatures of such a scenario, which—like signatures of supersymmetry or extra dimensions—would provide a major advance in our understanding of the fundamental structure of spacetime. There could be many consequences for particle physics, as in the well-studied Nelson--Barr models. In this paper, we have pointed out a key implication of such a scenario: it predicts that exactly stable domain walls were formed in the early universe, which must have been diluted by inflation. This opens up a new set of questions, motivated by quantum gravity, at the intersection between particle physics and cosmology.

Quantum gravity requires any would-be global symmetry to be either gauged or broken. While this is an old idea, it has recently been appreciated that the absence of \emph{all} global symmetries imposes dramatically more nontrivial constraints than may have been previously expected. For example, gauge theories come with a rich set of possible global symmetries (including generalized and non-invertible symmetries), which require intricate structure in quantum gravity to avoid. In this paper, we extended this perspective to include the generalized $\ZZ_2$ symmetries associated to Stiefel-Whitney classes of spacetime. These generalized symmetries arise not from an internal gauge theory, but from the gauging of parity, a spacetime symmetry. We saw qualitative differences between spacetime symmetries and internal symmetries. For example, the parity Wilson line operator, which counts parity domain walls modulo 2, is exactly topological up to the quantum gravity cutoff scale. It is not broken by any codimension-2 vortex. Nonetheless, we argued that there is not an associated global symmetry. The existence of end-of-the-world branes ultimately eliminates the topological parity Wilson line. In general, there are always dynamical cobordism defects to eliminate all potential topological operators. The Swampland Cobordism Conjecture gives a way to efficiently enumerate all possible topological operators or global symmetries, and to understand the sense in which they are gauged or broken. It sharpens and subsumes the ``No Global Symmetries'' hypothesis.

The modern perspective on generalized symmetries provides a new lens to reexamine many familiar ideas in physics and mathematics. Our results serve as an example of how this modern perspective, in the context of quantum gravity, can impose phenomenologically-relevant constraints on the cosmology of specific models of particle physics. We hope that such examples will encourage new interactions among disparate communities in theoretical physics, who have much to learn from each other.

\section*{Acknowledgments}

We thank Arun Debray, Daniel Harlow, Michael Hopkins, Miguel Montero, and Juven Wang for useful discussions.
We thank Daniel Aloni, Pouya Asadi, JiJi Fan, Sam Homiller, Qianshu Lu, John Stout, and Motoo Suzuki for discussions and feedback on an early draft.
MR is supported in part by the Alfred P.~Sloan Foundation Grant No.~G-2019-12504, the DOE Grant DE-SC0013607, and the NASA Grant 80NSSC20K0506. JM is supported by the U.S.~Department of Energy, Office of Science, Office of High Energy Physics, under Award Number DE-SC0011632.

\bibliography{ref}
\bibliographystyle{utphys}

\end{document}